\documentclass[a4paper,11pt]{article}
\usepackage{jheppub} 
\usepackage[utf8]{inputenc}
\usepackage{physics}
\usepackage{slashed}
\usepackage{xcolor}
\usepackage{comment}
\usepackage{multirow}
\usepackage{graphics}
\usepackage{float}
\usepackage{cancel}
\usepackage{soul}
\usepackage{cases}
\usepackage{array}
\usepackage{mathtools}  
\usepackage{amsfonts}
\usepackage{hyperref}
\usepackage{amsmath}
\usepackage{amssymb}
\usepackage{accents}
\usepackage{tcolorbox}
\usepackage{tikz}

\usepackage{amsmath,amssymb,bbm}
\usepackage{booktabs,longtable,array}

\usepackage[table]{xcolor}
\usepackage{tcolorbox}
\tcbuselibrary{skins, breakable}
\usepackage{float}
\newcommand{\ctil}[1]{\tilde{#1}}        

\newcommand{\Block}{\mathcal{K}}

\newcommand{\Vol}{\mathrm{Vol}}

\renewcommand{\braket}[1]{\langle0| #1 |0\rangle}

\newcommand{\R}{\mathbb{R}}

\newcommand{\Sgn}{\mathrm{Sgn}}

\newcolumntype{F}{>{\centering\arraybackslash$\displaystyle}m{3.8cm}<{$}}
\newcolumntype{R}{>{\centering\arraybackslash$\displaystyle}m{10.2cm}<{$}}

\title{$\mathbb{R}$eal Ambitwistors \& Massive Twistors for CFT$_4$}

\author{Aswini Bala,  Sachin Jain,  Deep Mazumdar, Adithya A Rao}

\affiliation{Indian Institute of Science Education and Research,\\ Dr Homi Bhabha Road, Pashan, Pune, India}

\emailAdd{aswini.bala@students.iiserpune.ac.in}
\emailAdd{sachin.jain@iiserpune.ac.in}
\emailAdd{deepkamal.mazumdar@students.iiserpune.ac.in}
\emailAdd{adithya.arao@students.iiserpune.ac.in}

\abstract{We develop a real twistor-space formulation of four-dimensional CFT Wightman correlators. We write the conformal generators and little-group constraints in ambitwistor space. The quadric condition $Z\cdot W$=0 arises naturally from the little-group constraints. We then solve for both the parity-even and parity-odd correlators in the ambitwistor variables. We further establish the connection between the ambitwistor correlators and the ones in the helicity-basis Grassmannian via a half-Fourier transform, and also develop the Penrose transform to recover the corresponding position-space correlators. We notice that for correlators with multiple tensor structures, the Penrose transform yields the full space of conformally invariant structures, but regularizing the associated Schwinger integrals selects the physical correlator. Finally, starting from the covariant Grassmannian formulation, we construct massive twistors that make the little-group symmetries manifest and develop their Penrose transform, which again reproduces the correct position-space correlators.}

\begin{document}

\maketitle

\section{Introduction}

Twistor theory was originally introduced by Penrose as a reformulation of spacetime physics \cite{Penrose:1967wn}, which relates the null lines in spacetime to points in twistor space and the lines in twistor space to points in spacetime, thereby replacing spacetime equations by geometric structures in twistor space. The modern utility of twistors was most vividly demonstrated in flat-space scattering amplitudes. Witten's twistor-string \cite{Witten:2003nn} proposal revealed that amplitudes possess a remarkably simple geometric organization in twistor space. This viewpoint led to developments in scattering amplitudes \cite{Berkovits:2004hg,Arkani-Hamed:2009hub,Mason:2009sa}, twistor strings \cite{Skinner:2013xp,Mason:2013sva,Geyer:2014fka,Gaberdiel:2021qbb}, Grassmannian formulations \cite{Arkani-Hamed:2012zlh}, and momentum twistors \cite{Mason:2009qx,Arkani-Hamed:2009nll}, exposing structures that are obscured in conventional formulations. 

Subsequently, twistors were developed for AdS spacetime, where twistors provide another way to represent bulk dynamics \cite{CarrilloGonzalez:2022ggn,Beetar:2024ptv,CarrilloGonzalez:2025qjk}: particularly for free fields and bulk propagators \cite{Adamo:2016rtr}, Wightman functions \cite{Ansari:2025fvi}, and ambitwistor string formulations \cite{Roehrig:2020kck}. However, the progress in AdS has been limited due to technical difficulties: the interaction vertices become non-local in twistor space, and the bulk-point integrals do not admit a simple twistor representation. As a result, existing constructions are largely restricted to propagators or special classes of observables. This makes it difficult to extend twistor methods to go beyond two-point functions, especially for spinning correlators.

A complementary approach is to use the boundary conformal invariance to alleviate some of these issues for boundary correlators. Twistors provide a natural language for conformal physics by furnishing a linear action of the conformal group and making the conservation trivial \cite{Baumann:2024ttn}, thus rendering a more economical description \cite{Baumann:2024ttn,Bala:2025gmz,Bala:2025jbh,Bala:2025qxr,Mazumdar:2025egx}.

In this work, we develop a twistor-space formulation of four-dimensional CFT Wightman correlators in Klein signature. We first define the real ambitwistors in Klein signature\footnote{The real ambitwistor was realized using the Grassmannian in \cite{Bala:2026trw}. See also \cite{CarrilloGonzalez:2026eum} for complex ambitwistors, which are defined for the complexified four-dimensional conformal group} and express the generators of the conformal group $SL(4,\mathbb R)$, along with the generators associated with the little group $GL(1,\R)\cross SL(2,\R)$ in terms of ambitwistors. Here, we obtain the quadric nature $Z\cdot W=0$ naturally as a consequence of little-group properties, instead of additionally imposing it. We then solve for the ambitwistor Wightman functions by using a Schwinger-parameterized ansatz, which manifestly incorporates conformal symmetry $SL(4,\mathbb R)$. We then solve for the coupled differential equations associated with the little group $GL(1,\R)\cross SL(2,\R)$, which results in two- and three-point Wightman functions of scalar operators and conserved currents, including the stress tensor. We obtain both: the parity-even and parity-odd ambitwistor correlators and observe that the total number of correlators obtained by this route for a particular spin configuration matches the independent counting obtained by demanding conservation and conformal invariance \cite{Stanev:2012nq, Costa:2011mg}.

We then derive the same correlators from a complementary perspective, taking inspiration from the recent developments in the Grassmannian formulation for CFT correlators \cite{Arundine:2026fbr,De:2026shn,Bala:2026hdm,Bala:2026bdx,Huang:2026tsh,Arundine:2026myr,Bala:2026lvw,Bala:2026trw}. Starting from the helicity-basis Grassmannian representation of momentum-space CFT$_4$ correlators \cite{Bala:2026trw}, we perform a half-Fourier transform to obtain their ambitwistor counterparts. In this correspondence, the Schwinger parameters appearing in the ambitwistor are naturally identified with those in the Grassmannian setup. We further develop the Penrose transform relating the twistor construction to position-space correlators. While the transform is unambiguous for correlators with a unique position-space tensor structure, such as $\langle OOO\rangle,\; \langle JOO\rangle,\; \langle TTO\rangle$, etc, it generates the full space of allowed conformally invariant tensor structures for other correlators like $\braket{JJJ}, \braket{TJJ}$, etc. We resolve this ambiguity by introducing an appropriate regularization of the Schwinger integrals, thereby reproducing the physical position-space correlators.

We finally present an alternate formulation of twistor-space correlators. Starting with the half-Fourier transform of the covariant Grassmannian correlators \cite{Bala:2026trw}, we obtain their counterparts in terms of massive twistors. A crucial feature of these massive twistors is that they make the $SL(2,\mathbb{R})$ little-group symmetry manifest. We then develop the Penrose transform for the correlators in massive twistor variables, and obtain the correct position-space correlators via this transform. 

This establishes a twistor-space representation of CFT$_4$ correlators and connects it to both the Grassmannian setup of momentum-space and the position-space formulations.

The paper is organized as follows: In \textit{section} \ref{sec:AmbiTwistors}, we start by developing the ambitwistor variables and deriving the conformal generators and the little-group constraints. We then solve for the two- and three-point correlation functions for $\Delta=2$ scalars and conserved currents using these constraints for both the parity-even and parity-odd correlators. Furthermore, we establish a connection between these ambitwistor correlators and those in the helicity-basis Grassmannian via a half-Fourier transform. We finally conclude this section by deriving the Penrose transform, which takes these ambitwistor correlators to their correct position-space counterparts. In \textit{section} \ref{sec:SL2RTwistors}, we start with the correlators in the covariant Grassmannian and perform a half-Fourier transform to obtain their analogs in the massive-twistor variables, which enjoy manifest little-group covariance. We then develop the Penrose transform for massive twistors as well, which again results in the correct position-space correlators. Finally, in \textit{section} \ref{sec:Disc}, we summarize this work and discuss a number of interesting future directions.

We supplement this paper with a few essential appendices. We present a brief review of four-dimensional off-shell spinor-helicity in \textit{appendix} \ref{app:OffshellSH}. We then present the details of solving the little-group constraints in \textit{appendix} \ref{app:JOO-fc-derive}. In \textit{appendix} \ref{app:AmbiDetail}, we give details of the necessary Schwinger integrals involved in ambitwistor correlators. We provide the detailed calculation of the Penrose transform for ambitwistor correlators in \textit{appendix} \ref{app:penrose-details}. Lastly, we present the bulk computation of Chern-Simons theory in \textit{appendix} \ref{CST}.

\section{Ambitwistors for CFT\(_4\) Correlators}\label{sec:AmbiTwistors}
The conformal group in four dimensions is isomorphic to the Lorentz group in six dimensions. We choose to work in the Klein signature to obtain a real representation of the twistor space. In $(2,2)$ signature, the conformal algebra is given by {${conf}(2,2) \simeq spin(3,3)\simeq SL(4,\mathbb{R})$}. In ambitwistor space, we have a pair of twistor $Z$ and dual twistor $W$, which are the fundamental and anti-fundamental representations of the conformal group.

We start with the spinor-helicity representation of four-dimensional off-shell momenta, 
\begin{align}\label{AmbiSH}
p_{\alpha\dot{\alpha}}=\lambda_{I\alpha}\epsilon^{IJ} \tilde\lambda_{J\dot\alpha} \equiv \lambda_\alpha \tilde\rho_{\dot\alpha}-\rho_\alpha \tilde\lambda_{\dot\alpha},
\end{align}
and perform a half-Fourier transform over $\rho_\alpha,\;\tilde\rho_{\dot\alpha}$ to obtain a pair of twistors,\footnote{Our version of ambitwistor differs from the one in \cite{CarrilloGonzalez:2026eum}, in the sense that our ambitwistors are in Klein-space, while theirs are in complexified Minkowski-space.} 
\begin{align}\label{Ambitwistor}
(\lambda_\alpha,\tilde{\rho}_{\dot{\alpha}} )\;\xrightarrow{HFT}&\;Z^A=(\lambda_\alpha,\tilde{\mu}^{\dot{\alpha}} ),\notag\\
(\tilde{\lambda}_{\dot{\alpha}},\rho_\alpha )\;\xrightarrow{HFT}&\;W_A=(\mu^{\alpha},\tilde{\lambda}_{\dot{\alpha}} ).
\end{align}
Notice that here \(W\) is not the Fourier conjugate of \(Z\) but is rather independent of \(Z\).\\
Our objects of interest here are the conserved currents in the helicity basis. In order to get those, one contracts the currents with the polarization vectors of appropriate helicity, eq \eqref{eq:Helicity-basis-spinors}\footnote{Refer to appendix \ref{app:OffshellSH} for details on construction of off-shell spinor helicity variables.}. 
\begin{equation}
\xi^{(+)}_{\alpha\dot\alpha}
=\frac{\lambda_\alpha\tilde{\lambda}_{\dot\alpha}}{p},
\qquad
\xi^{(-)}_{\alpha\dot\alpha}
=\frac{\rho_\alpha\tilde{\rho}_{\dot\alpha}}{p},
\qquad
\xi^{(0)}_{\alpha\dot\alpha}
=\frac{\rho_\alpha\tilde{\lambda}_{\dot\alpha}+\lambda_\alpha\tilde{\rho}_{\dot\alpha}}{p}.\label{eq:Helicity-basis-spinors-mt}
\end{equation}
The momentum bispinor written in the above form, eq \eqref{AmbiSH}, has a little group redundancy of \(SL(2,\mathbb{R})\times GL(1,\mathbb{R})\). Under the action of the little group generators, these polarization vectors also transform covariantly; therefore, the conserved current written in the helicity basis will also transform covariantly under little group transformations.
Consequently, the correlator must satisfy not only the conformal Ward identities but also the little-group covariance. The construction of the conformal and little-group generators will be the subject of the next section.

\subsection{Conformal Generators and Little Group}\label{Generators}
 
The differential generators of the $SL(4,\mathbb{R})$ in the ambitwistor setup is given by
\begin{align}\label{AmbiGen}
T^{A}_B = Z^A \frac{\partial}{\partial Z^B} - W_B\frac{\partial}{\partial W_A}   - \frac{1}{4}\delta^{A}_B \left(  Z^C \frac{\partial}{\partial Z^C} - W_C\frac{\partial}{\partial W_C}   \right),
\end{align}
where $A,\; B$ and \(C\) are the indices of the fundamental representation of $SL(4,\mathbb{R})$. The generators eq \eqref{AmbiGen} satisfy the $SL(4,\mathbb{R})$ algebra
\begin{align}
\big[ T^A_B, T^C_D\big] = \delta_B^CT_D^A-\delta_D^AT_B^C.
\end{align} 
The differential generator of \(GL(1,\mathbb{R})\) takes the following form
\begin{align}\label{AmbiLittleG}
G&=\frac{1}{2}\Big(Z^A\frac{\partial}{\partial Z^A}-W_A\frac{\partial}{\partial W_A}\Big).
\end{align}
and we demand the following action of \(GL(1,\mathbb{R})\) generators on the ambitwistor operators,\footnote{This is because the polarization vectors for integer spin currents eq \eqref{eq:Helicity-basis-spinors} are annihilated by \(GL(1,\mathbb{R})\) transformations, but the polarization spinors for half-integer spin currents eq \eqref{Helicity-basis-polarization-spinors} give an eigenvalue \(\pm s\).}
\begin{equation}
    G[\tilde J_s^{\pm ,0} (Z,W)] = \begin{cases}
    0 ~& s\in \mathbb{Z}_{\ge 0}\\
    \pm \, \tilde J^{\pm,0}_s(Z,W) ~&s\in \mathbb{Z}_{\ge 0} + \frac{1}{2} 
\end{cases}
\end{equation}
where \(s\) denotes the spin and \(\pm,0\) represent the helicity of the operator. \\
The \(SL(2,\mathbb{R})\) generators take on the following form in the twistor space
\begin{align}\label{AmbiLittleS}
S^{I}_{J} = \begin{bmatrix}\frac{1}{2}\left( Z \cdot \frac{\partial}{\partial Z}+ W \cdot \frac{\partial}{\partial W}\right) + 2 & i (Z \cdot W) \\
-i \frac{\partial^2}{\partial Z \cdot \partial W} & -\frac{1}{2}\left( Z \cdot \frac{\partial}{\partial Z}+ W \cdot \frac{\partial}{\partial W}\right) - 2 \end{bmatrix},
\end{align}
We demand the following eigenvalue equation of the $SL(2, \mathbb{R})$ diagonal generator for the ambitwistor operators
\begin{align}\label{AmbiDiag}
{S^{(1)}_{(1)} \left[ \tilde{J}^{\pm,0}_s(Z, W) \right]=-S^{(2)}_{(2)} \left[ \tilde{J}_s^{\pm,0}(Z, W) \right] = \left( \frac{+}{0} \right) s \left[ \tilde{J}_s^ {\pm,0}(Z, W) \right]},
\end{align}
The off-diagonal pieces, namely \(S^1_2\) and \(S^2_1\), are helicity-raising and lowering operators, respectively, and we demand that the correlator of highest helicity be annihilated by the raising operator and lowest by the lowering operator.

With these generators in hand, we solve for the correlation functions in the ambitwistor setup using the conformal Ward identities eq \eqref{AmbiGen}, along with the little group operators eq \eqref{AmbiLittleG} and eq \eqref{AmbiLittleS}. This will be the subject of our discussion in the next section.

\subsection{Conformal Correlators From Solving Ward identities}\label{Correlators}


Any function that depends only on \(SL(4,\mathbb{R})\) invariant objects is a solution to the conformal Ward identity eq \eqref{AmbiGen}. Since \(SL(4,\mathbb{R})\) is not equipped with an invariant metric, the only invariant object that can be constructed out of the twistor ($Z$) and the dual twistor ($W$) is \(Z_i\cdot W_j \equiv Z_i^A W_{jA}\).\footnote{There are also distributional solutions which are schematically of the form \(\int dC\, \delta(C\cdot Z)\) \& \(\int d\tilde C\, \delta(\tilde C\cdot W)\) but we do not consider these in our work. Beyond three points, there are other invariants that can be constructed from the Levi-Civita symbol, but for our discussion, this invariant is irrelevant.} Taking inspiration from CFT\(_3\) \cite{Baumann:2024ttn,Bala:2025gmz,Bala:2025jbh,Bala:2025qxr,Mazumdar:2025egx} correlators and also anticipating a connection with the CFT\(_4\) Grassmannian \cite{Bala:2026trw}, we consider here the following general solution, written completely in terms of the Schwinger parameters
\begin{align}\label{ansatz}
\boxed{\braket{J_{s_1}^{h_1}(Z_1, W_1) \cdots J_{s_n}^{h_n}(Z_n, W_n)}  =\int d^{n^2}c_{mn} \,
  e^{ic_{ij}Z_{i} \cdot W_j}\mathcal{G}^{h_1\cdots h_n}_{s_1\cdots s_n}(c_{mn}),}
\end{align}
where $\mathcal{G}^{h_1\cdots h_n}_{s_1\cdots s_n}(c_{mn})$ is a function of the Schwinger parameters $c_{ij}$, which depends on the spin \(s\) and helicity \(h\) of the operators in the particular correlator. 
The function \(\mathcal{G}^{h_1\cdots h_n}_{s_1\cdots s_n}\) can be fixed by imposing little-group covariance on the twistor correlator. In particular, we act with the little-group generators eq \eqref{AmbiLittleG} and eq \eqref{AmbiLittleS} on the general ansatz eq \eqref{ansatz} and use integration by parts to express their action entirely in terms of the Schwinger parameters. The resulting \(GL(1,\mathbb{R})\) constraints take the form
\begin{align}\label{AmbiGL1}
\sum_{j} \left(\frac{\partial}{\partial c_{kj}} -\frac{\partial}{\partial\, c_{jk}}\right)\mathcal{G}^{h_1\cdots h_n}_{s_1\cdots s_n}(c_{mn}) =\begin{cases}
    0 ~& s\in \mathbb{Z}_{\ge 0}\\
   -2 h_k\mathcal{G}^{h_1\cdots h_n}_{s_1\cdots s_n}(c_{mn}) ~&s\in \mathbb{Z}_{\ge 0} + \frac{1}{2} 
\end{cases},~~\forall\; k.
\end{align}
The diagonal part of the \(SL(2,\mathbb{R})\) leads to the following helicity eigenvalue equation 
\begin{align}\label{AmbiSL2Diag}
 \sum_{j} \left(\frac{\partial}{\partial c_{kj}} (c_{kj} \mathcal{G}^{h_1\cdots h_n}_{s_1\cdots s_n}(c_{mn})) +\frac{\partial}{\partial c_{jk}} (c_{jk} \mathcal{G}^{h_1\cdots h_n}_{s_1\cdots s_n}(c_{mn}))\right) = (4-2h_k )\mathcal{G}^{h_1\cdots h_n}_{s_1\cdots s_n}(c_{mn}),~~\forall\; k.
\end{align}
while the off-diagonal helicity-raising and helicity-lowering operators impose the following constraints, respectively
\begin{align}\label{AmbiSL2Raise}
\frac{\partial}{\partial c_{kk}} \mathcal{G}^{h_1\cdots h_k\cdots h_n}_{s_1\cdots s_n}(c_{mn}) = \mathcal{G}^{h_1\cdots (h_k+1)\cdots h_n}_{s_1\cdots s_n}(c_{mn}), 
\end{align}
\begin{align}\label{AmbiSL2Lower}
-\sum_{i,j} \frac{\partial}{\partial c_{ij}}(c_{ik}c_{kj}\, \mathcal{G}^{h_1\cdots h_k\cdots h_n}_{s_1\cdots s_n}(c_{mn})) ~+~4c_{kk}\mathcal{G}^{h_1\cdots h_k\cdots h_n}_{s_1\cdots s_n}(c_{mn}) = \mathcal{G}^{h_1\cdots (h_k-1)\cdots h_n}_{s_1\cdots s_n}(c_{mn}).
\end{align}

With the general ansatz eq \eqref{ansatz} and the little-group operators eq \eqref{AmbiGL1}-\eqref{AmbiSL2Lower} at our disposal, we now solve for various ambitwistor correlators, starting with the two-point functions.

\subsubsection*{Two-Point Functions}
We start with the simple example of the scalar two-point function $\braket{ O_2O_2}$.\\
\(\mathbf{\braket{O_2 O_2}}\)\\
 The \(GL(1,\R)\) constraint eq \eqref{AmbiGL1} gives 
\begin{align}
c_{21}\partial_{c_{21}}\mathcal{G}_{OO} - c_{12}\partial_{c_{12}}\mathcal{G}_{OO} = 0.
\end{align}
The \(SL(2,\R)\) raising constraint eq \eqref{AmbiSL2Raise} gives $\mathcal{G}_{OO}$ is independent of \(c_{11}\) and \(c_{22}\), while the \(SL(2,\R)\) diagonal constraint eq \eqref{AmbiDiag} results in
\begin{equation}
    c_{21}\partial_{c_{21}}\mathcal{G}_{OO} + c_{12}\partial_{c_{12}} \mathcal{G}_{OO} = 0.
\end{equation} 
It is immediately clear that the above constraints are satisfied only if \(\mathcal{G}_{OO}\) is a constant. Therefore, the twistor-space expression for \(\braket{O_2 O_2}\) is simply
\begin{equation}
   {\braket{O_2 O_2} = \int dc_{mn}\,\exp(i c_{ij}Z_i\cdot W_j)\, 1 = \delta(Z_1\cdot W_1)\delta(Z_2\cdot W_2)\delta(Z_1\cdot W_2)\delta(Z_2\cdot W_1).}\label{eq:o2o2-ambitwistor}
\end{equation}
The appearance of the constraint \(\delta(Z_i\cdot W_j)\) shows that the correlator is supported on the quadric \(Z\cdot W = 0\), which is part of the usual definition of the ambitwistor space. These delta functions (and their derivatives, which will appear in other correlators) are of utmost importance in making a connection to the momentum-space via the Grassmannian, which we will show in section \ref{Grasstotwistor}. Later, when performing the Penrose transform, we will also see that the deltas and their derivatives do not contribute any tensor structures, and therefore act as trivial constraints under the Penrose transform.\\


\noindent\(\mathbf{\braket{J_s J_s}}\)\\
We now consider the spinning case, starting with conserved spin-1 currents in all-plus helicity configuration. The \(GL(1,\R)\) constraint once again gives 
\begin{equation}
c_{21}\partial_{c_{21}}\mathcal{G}_{J^+J^+} - c_{12}\partial_{c_{12}}\mathcal{G}_{J^+J^+} = 0,
\end{equation}
while the \(SL(2,\R)\) raising constraint again imposes that $\mathcal{G}_{J^+ J^+}$ is independent of \(c_{11}\) and \(c_{22}\). The diagonal constraint of \(SL(2,\R)\) for spin-1 current results in
\begin{equation}
c_{21}\partial_{c_{21}}\mathcal{G}_{J^+J^+} + c_{12}\partial_{c_{12}}\mathcal{G}_{J^+J^+} = -2\mathcal{G}_{J^+J^+}.
\end{equation}
The unique solution to this set of differential equations is 
\begin{equation}
\mathcal{G}_{J^+J^+} = \frac{K}{c_{12}c_{21}},
\end{equation}
and the twistor-space correlator is therefore 
\begin{align}
    {
    \begin{aligned}
        \braket{J^+J^+} &=\int dc_{mn}\,\exp(i c_{ij}Z_i\cdot W_j)\, \frac{1}{c_{12}c_{21}}\\
        &= \delta(Z_1\cdot W_1)\delta(Z_2\cdot W_2)\,\Sgn(Z_1\cdot W_2)\,\Sgn(Z_2\cdot W_1),
    \end{aligned}
    }\label{eq:Ambi-JpJp}
\end{align}
while the results for other helicities can be obtained via the helicity-lowering operator. Similarly, the general integer spin-s correlator in the all-plus helicity configuration is simply
\begin{align}
{\mathcal{G}_{J_s^+J_s^+} = \frac{K}{(c_{12}c_{21})^s}.}
\end{align}
Specifically, the two-point stress-tensor correlator takes on the following form
\begin{equation}
\boxed
{\braket{T^+T^+} =\int dc_{mn}\,\exp(i c_{ij}Z_i\cdot W_j)\, \frac{1}{c_{12}^2c_{21}^2} = \delta(Z_1\cdot W_1) \delta(Z_2\cdot W_2) |Z_1\cdot W_2| |Z_2\cdot W_1|.}
\end{equation}\\

\noindent\(\mathbf{\braket{\psi_{\frac{1}{2}} \bar\psi_{\frac{1}{2}}}}\)\\
To obtain two-point correlators of fermionic operators, one would take a similar route, but now by demanding the eigenvalue \(h_k\) under the \(GL(1,\mathbb{R})\) constraint eq \eqref{AmbiGL1}. With this modification, one can again easily solve for the correlator, and the two-point fermionic correlator in all-plus helicity takes on the form 
\begin{equation}
    {\braket{\psi^+_\frac{1}{2} \bar \psi^+_\frac{1}{2}} =\int dc_{mn}\,\exp(i c_{ij}Z_i\cdot W_j)\, \frac{1}{c_{12}} =   \delta(Z_1\cdot W_1)\delta(Z_2\cdot W_2)\delta(Z_2\cdot W_1) \text{Sgn}(Z_1\cdot W_2).}
    \label{eq:ambi-psiplus}
\end{equation}
while in all-minus helicity, the correlator takes on the form 
\begin{align}
    \notag \braket{\psi^-_\frac{1}{2} \bar \psi^-_\frac{1}{2}} &=\int dc_{mn}\,\exp(i c_{ij}Z_i\cdot W_j)\,\left( \frac{c_{11} c_{22}}{c_{12}}-c_{21}\right) \\
    &=  \delta^{[1]}(Z_1\cdot W_1) \delta^{[1]}(Z_2\cdot W_2) \text{Sgn}(Z_1\cdot W_2) \delta(Z_2\cdot W_1)  \notag \\
    &~~~~~~~~~~~~~~~~~~~~~~~~-\delta(Z_1\cdot W_1) \delta(Z_2\cdot W_2)\delta(Z_1\cdot W_2)\delta^{[1]}(Z_2\cdot W_1).
\end{align}
The appearance of \(\delta^{[1]}(Z_i\cdot W_i)\) is very crucial for the correlator to have the right properties under the \(SL(2,\mathbb{R})\) raising and lowering operators. In twistor space, the raising operator is \(Z_i\cdot W_i\), and acting this on the \(\braket{\psi^-_{\frac{1}{2}}\bar\psi^-_{\frac{1}{2}}}\) correlator, we get 
\begin{align}
    \notag (Z_1\cdot W_1)(Z_2\cdot W_2)&\braket{\psi^-_\frac{1}{2}\bar\psi^-_\frac{1}{2}}\\= &(Z_1\cdot W_1) (Z_2 \cdot W_2)\Big( \delta^{[1]}(Z_1\cdot W_1) \delta^{[1]}(Z_2\cdot W_2) \text{Sgn}(Z_1\cdot W_2) \delta(Z_2\cdot W_1)\notag \\
    \notag &~~~~~~~~~~~~~~~~~~~~~~~~~~~~-\delta(Z_1\cdot W_1) \delta(Z_2\cdot W_2)\delta(Z_1\cdot W_2)\delta^{[1]}(Z_2\cdot W_1) \Big)\\
     \notag = & \delta(Z_1\cdot W_1)\delta(Z_2\cdot W_2)\delta(Z_2\cdot W_1) \text{Sgn}(Z_1\cdot W_2)\\
     = & \braket{\psi^+_\frac{1}{2}\bar\psi^+_\frac{1}{2}}.
\end{align}
If the \(\braket{\psi^-_\frac{1}{2}\bar\psi^-_\frac{1}{2}}\) correlator did not have the appropriate delta functions and their derivatives, then it would not have been possible to go from the all-minus to the all-plus correlator using the helicity-raising operator. Also note that the all-plus helicity correlator does not have derivatives of \(\delta(Z_i\cdot W_i)\), and therefore is annihilated by the raising operator, while the all-minus correlator has the correct structure such that it is annihilated by the lowering operator eq \eqref{AmbiLittleS}. Thus, the distributional structure involving \(\delta(Z_i\cdot W_i)\) and its derivatives is essential for realizing the full \(SL(2,\mathbb{R})\) little-group representation in twistor space.

We now move on to the three-point functions.

\subsubsection*{Three-Point Functions in Schwinger Parameter Space}

The three-point functions can be similarly solved for by using the little group constraints. Refer to the appendix \ref{app:JOO-fc-derive} for a detailed calculation. Here we present only the results of solving for \(\mathcal{G}(c_{mn})\) for some correlators

\begin{table}[h]
    \centering
    \renewcommand{\arraystretch}{2.07}
    \begin{tabular}{|c|l|}
        \hline
        Correlator & $\mathcal{G}(c_{mn})$ \\
        \hline
        \(\braket{O_2 O_2 O_2}\) & \(\displaystyle\frac{1}{(c_{12}c_{23}c_{31} - c_{13} c_{32} c_{21})}\)\\
        \(\braket{J^+ O_2 O_2}\) & \(\displaystyle\frac{c_{23}c_{32}}{(c_{12}c_{23}c_{31} - c_{13} c_{32} c_{21})^2}\)\\
        \(\braket{J^+ J^+ O_2}\) & \(\displaystyle\frac{c_{13}c_{23}c_{32}c_{31}}{(c_{12}c_{23}c_{31} - c_{13} c_{32} c_{21})^3}\)\\
        \(\braket{T^+T^+O_2}\) & \(\displaystyle \frac{c_{13}^2 c_{23}^2 c_{32}^2 c_{31}^2}{\left(c_{12} c_{23} c_{31}-c_{13} c_{21} c_{32}\right){}^5}\)\\
        \(\braket{J^+J^+J^+}\) & \(\displaystyle\mathcal{C}_1 \frac{1}{ \left(c_{12} c_{23} c_{31}-c_{13} c_{21} c_{32}\right){}^2} +\,\mathcal{C}_2 \frac{ c_{12} c_{13} c_{21} c_{23} c_{32} c_{31}}{\left(c_{12} c_{23} c_{31}-c_{13} c_{21} c_{32}\right){}^4}\) \\
        & \(\displaystyle +\, \mathcal{D}_1 \frac{\left(c_{12} c_{23} c_{31}+c_{13} c_{21} c_{32}\right)}{\left(c_{12} c_{23} c_{31}-c_{13} c_{21} c_{32}\right){}^3}\)\\
        \(\braket{T^+J^+J^+}\)  & \(\mathcal{C}_1 \displaystyle \frac{c_{23} c_{32}}{ \left(c_{12} c_{23} c_{31}-c_{13} c_{21} c_{32}\right){}^3} + \displaystyle\mathcal{C}_2 \frac{c_{23}^2 c_{32}^2 c_{12} c_{21} c_{13} c_{31}}{\left(c_{12} c_{23} c_{31}-c_{13} c_{21} c_{32}\right){}^5}\)\\
        & \(\displaystyle +\mathcal{D}_1 \frac{c_{23} c_{32} \left(c_{12} c_{23} c_{31}+c_{13} c_{21} c_{32}\right) }{\left(c_{12} c_{23} c_{31}-c_{13} c_{21} c_{32}\right){}^4}\)\\
        \(\braket{T^+T^+T^+}\) & \(\displaystyle\mathcal{C}_1 \frac{1}{\left(c_{13} c_{21} c_{32}-c_{12} c_{23} c_{31}\right){}^3} + \displaystyle\mathcal{C}_2 \frac{ c_{12} c_{23} c_{31}c_{13} c_{21} c_{32}}{ \left(c_{12} c_{23} c_{31}-c_{13} c_{21} c_{32}\right){}^5}\)\\
        & \(\displaystyle +\, \mathcal{C}_3 \frac{\left(c_{12}c_{13}c_{21}c_{23}c_{31}c_{32}\right)^2}{\left(c_{12} c_{23} c_{31}-c_{13} c_{21} c_{32}\right){}^7} + \mathcal{D}_1 \frac{(c_{12} c_{23} c_{31}+c_{13} c_{21} c_{32})}{ \left(c_{12} c_{23} c_{31}-c_{13} c_{21} c_{32}\right){}^4} \)\\
        & \(\displaystyle +\, \displaystyle \mathcal{D}_2 \frac{c_{12} c_{13} c_{21} c_{23} c_{31} c_{32} \left(c_{12} c_{23} c_{31}+c_{13} c_{21} c_{32}\right)}{ \left(c_{12} c_{23} c_{31}-c_{13} c_{21} c_{32}\right){}^6}\) \\[0.3cm]
        \hline
    \end{tabular}
    \caption{Few examples of \(\mathcal{G}(c_{mn})\) (with the \(\Sgn\) function stripped off) obtained by solving for the little group constraints.}
    \label{tab:correlators}
\end{table}
An important observation is that if we plug the \(\mathcal{G}(c_{mn})\) corresponding to the correlator \(\braket{O_2 O_2 O_2}\) in the general ansatz eq \eqref{ansatz}, we obtain
\begin{equation}
    \braket{O_2(Z_1, W_1)O_2(Z_2, W_2)O_2(Z_3, W_3)} = \int d^{n^2}c_{mn} \,
  e^{ic_{ij}Z_{i} \cdot W_j}\frac{1}{c_{12}c_{23}c_{31} - c_{13}c_{32}c_{21}}.
\end{equation}
Under the exchange of any two legs, say \(1\leftrightarrow 2\), the exponential factor becomes
\begin{equation}
e^{ic_{ij}Z_i\cdot W_j}
\longrightarrow
e^{i\left(c_{12}Z_2\cdot W_1+c_{21}Z_1\cdot W_2+c_{13}Z_2\cdot W_3+c_{23}Z_1\cdot W_3+c_{31}Z_3\cdot W_2+c_{32}Z_3\cdot W_1
\right)}.
\end{equation}
We can restore the exponential to its original form by simultaneously relabeling the integration variables according to
\begin{equation}
c_{1k}\leftrightarrow c_{2k},
\qquad
c_{k1}\leftrightarrow c_{k2},
\qquad \forall, k.
\end{equation}
Under this relabeling, the integration measure is invariant, while the combination appearing in the denominator transforms as
\begin{equation}
c_{12}c_{23}c_{31}-c_{13}c_{32}c_{21} \longrightarrow -\left(c_{12}c_{23}c_{31}-c_{13}c_{32}c_{21}\right),
\end{equation}
and therefore, the correlator picks a minus sign under \(1\leftrightarrow 2\) exchange, which is not expected for a bosonic three-point correlator. Therefore, the full function \(\mathcal{G}(c_{mn})\) must also contain the factor
\begin{equation}
\Sgn\left(c_{12}c_{23}c_{31}-c_{13}c_{32}c_{21}\right),
\end{equation}
to ensure correct statistics. This is a general feature for all correlators, where each correlator must carry a \(\Sgn\) function to ensure the correct exchange (anti)symmetry.\\

Notice that the structures appearing in the \(\braket{JJJ}\), \(\braket{TJJ}\), and \(\braket{TTT}\) correlators in table \ref{tab:correlators} naturally fall into two classes. The structures multiplying \(\mathcal{C}_i\) have the exchange symmetry required by the external operators, while those multiplying \(\mathcal{D}_i\) have the opposite exchange behavior. With the notion of parity in twistor space, which we will discuss in detail in section \ref{sec:parity-odd}, we will see that the structures multiplying \(\mathcal{C}_i\) correspond to parity-even correlators, while those multiplying \(\mathcal{D}_i\) correspond to parity-odd correlators. 

These two classes account for the complete set of structures obtained from the little group constraints. In particular, we find exactly \(2\min(s_1,s_2,s_3)+1\) independent structures, with \(\min(s_1,s_2,s_3)+1\) parity-even structures multiplying \(\mathcal{C}_i\), and \(\min(s_1,s_2,s_3)\) parity-odd structures multiplying \(\mathcal{D}_i\). This precisely matches the expected counting of three-point structures\footnote{We have verified this up to \(s=5\).} \cite{Stanev:2012nq,Costa:2011mg}. This agreement provides a non-trivial consistency check of our construction and shows that the little group constraints capture the complete set of allowed three-point correlator structures.

\subsubsection*{Three-point correlators in twistor space}
Given the \(\mathcal{G}(c_{mn})\) in table \ref{tab:correlators}, we can convert them to twistor space by plugging them into the general ansatz eq \eqref{ansatz} and doing the \(c_{ij}\) integrals. Doing so, in twistor space, the scalar correlator $\langle O_2O_2O_2\rangle$ has the following form
\begin{align}
{
\begin{aligned}
    \braket{O_2O_2O_2} &= \int dc_{ij} ~\exp( i c_{ij} Z_i\cdot W_j) \frac{1}{|c_{12}c_{23}c_{31}-c_{13}c_{21}c_{32}|} \\ 
&=  \frac{\delta(Z_1\!\cdot\! W_1)\,
\delta(Z_2\!\cdot\! W_2)\,
\delta(Z_3\!\cdot\! W_3)}
{|(Z_1\!\cdot\! W_2)\,
(Z_2\!\cdot\! W_3)\,
(Z_3\!\cdot\! W_1)|} ~ \log(Z_1\cdot W_2\, Z_2\cdot W_3\, Z_3\cdot W_1) \delta(1+\tau),
\end{aligned}
}
\label{eq:Ambi-o2o2o2}
\end{align}
where \(\tau\) is a three-point cross-ratio given by 
\begin{equation}
    \tau = \frac{Z_1 \cdot W_3~Z_3\cdot W_2~Z_2\cdot W_1}{Z_1 \cdot W_2~Z_2\cdot W_3~Z_3\cdot W_1}.    
\end{equation}
The way to perform the $c_{ij}$ integral in eq \eqref{eq:Ambi-o2o2o2} has been explicitly worked out in the appendix \ref{app:AmbiDetail}. We would like to mention here that the cross ratio \(\tau\) also appears separately in the twistor-space analysis in \cite{Bala:2026trw,CarrilloGonzalez:2026eum}, and plays a non-trivial role in the calculations in \cite{CarrilloGonzalez:2026eum}. On the other hand, our twistor-space answers cannot have a non-trivial dependence on \(\tau\) since at the support of the delta function we have \(\tau = -1\). 

\subsubsection*{Spinning Three-Point Correlators}

Converting the \(\braket{J^+O_2O_2}\) correlator to twistor space, we get 
\begin{align}
{
\begin{aligned}
    \langle J^+ O_2 O_2\rangle
    &= \int dc_{ij} ~\exp(ic_{ij} Z_i\cdot W_j) \frac{c_{23} c_{32}~ \text{Sgn}(c_{12}c_{23}c_{31}-c_{13}c_{21}c_{32})}{(c_{12}c_{23}c_{31}-c_{13}c_{21}c_{32})^2}\\ 
&=\delta(Z_1\!\cdot\! W_1)\,
\delta(Z_2\!\cdot\! W_2)\,
\delta(Z_3\!\cdot\! W_3)\frac{ \text{Sgn}\left(Z_1\cdot W_2\, Z_2\cdot W_3\, Z_3\cdot W_1\right) }
{(Z_2\!\cdot\! W_3)\,
(Z_3\!\cdot\! W_2)}\\
&\hspace{145pt}\log(Z_1\cdot W_2\, Z_2\cdot W_3\, Z_3\cdot W_1) \delta(\tau +1).
\end{aligned}
}
\end{align}
Similarly, we obtain the following expression for $\braket{J^+ J^+ O_2}$

\begin{align}
\label{eq:Ambitwistor-JJO}
{
\begin{aligned}
    \langle0|J^+J^+O_2|0\rangle &= \int \prod_{i,j=1}^3 dc_{ij} ~e^{i\sum_{i,j=1}^{3} c_{ij} Z_i\cdot W_j} \frac{c_{13}c_{23}c_{31}c_{32}}{|c_{12}c_{23}c_{31}-c_{13}c_{21}c_{32}|^3}  \\ 
&=
\delta(Z_1\!\cdot\! W_1)\,
\delta(Z_2\!\cdot\! W_2)\,
\delta(Z_3\!\cdot\! W_3)
\frac{
(Z_1\!\cdot\! W_2)(Z_2\!\cdot\! W_1)\,
}{
|(Z_1\!\cdot\! W_2)(Z_2\!\cdot\! W_3)
(Z_3\!\cdot\! W_1)|
}\\
&\hspace{130pt}\log(Z_1\cdot W_2\, Z_2\cdot W_3 \, Z_3\cdot W_1) \delta(\tau +1).
\end{aligned}
}
\end{align}
  \\
It is interesting to note that \(\delta(1+\tau)\) is present in all the correlators. The physical interpretation and significance of this constraint are not clear at present, and we defer a more detailed investigation to future work. 

Similarly, one can also get all spinning correlators like \(\braket{JJJ}\) in ambitwistor space. However, these correlators will have a closed-form expression only after using some scheme-dependent regularization, which gives an expression involving dilogs.

Correlators written in terms of the helicity-basis Grassmannian in \cite{Bala:2026trw} have a beautiful connection to the correlators we obtain in twistor space in terms of the Schwinger parameters, and we discuss this connection in detail in the next section.



\subsection{From Helicity Basis Grassmannian to Ambitwistors}\label{Grasstotwistor}
We start with the Grassmannian representation of a general correlator \cite{Bala:2026trw}
\begin{equation}\label{eq:grassmannian-correlator}
\psi_n^{h_1,\cdots, h_n}(\lambda, \tilde \lambda, \rho, \tilde\rho)=
 \int
 \frac{DC\,D\widetilde C}
 {\mathrm{Vol}\,GL(n)\,\mathrm{Vol}\,GL(n)}
 \delta(C\widetilde\Lambda)
 \delta(\widetilde C\Lambda)
 \delta(\widetilde C\Omega C^T)
 \mathcal{F}_n^{h_1,\cdots, h_n}(C,\widetilde C),
\end{equation}
where \(\Lambda = \{\lambda_1,\cdots, \lambda_n, \rho_1,\cdots, \rho_n\}\) and \(\tilde\Lambda = \{\tilde\lambda_1,\cdots, \tilde\lambda_n, \tilde\rho_1,\cdots \tilde\rho_n\}\). \\
We can gauge fix the \(GL(n)\) redundancy by choosing the \(\widetilde C_{n\times 2n}\) matrix to be of the form 
\begin{equation}
    \widetilde C = ( -c_{n\times n} |I_{n \times n}).
\end{equation}
The symplectic orthogonality  \(\widetilde C\Omega C^T =0\) then fixes the form of the \(C\) matrix to be \(C = (-c^T_{n\times n}|I_{n\times n})\), and we obtain the following expression for momentum-space correlators
\begin{equation}
    \psi_n^{h_1,\cdots, h_n}(\lambda, \tilde \lambda, \rho, \tilde\rho)= \int dc_{mn}  ~\delta(\tilde \rho_i - c_{ji}\tilde \lambda_j ) \delta(\rho_i - c_{ij} \lambda_j ) \mathcal{F}_n^{h_1,\cdots, h_n}(c_{mn}),
\end{equation}
where 
\begin{equation}
    \mathcal{F}_n(c_{mn}) = \mathcal{F}_n\left( (-c_{n\times n}^T~|~I_{n\times n}),~(-c_{n\times n}~|~I_{n\times n})  \right).
    \label{eq:ambi-gaugefix}
\end{equation}
Proceeding to perform the half-Fourier transform of the \(\rho_i\) and \(\tilde \rho_i\) variables, we have
\begin{align}
\notag\hat{\psi}_n^{h_1,\cdots, h_n}(Z, W) &=\int dc_{mn} d^2\tilde\rho_i~ d^2\rho_i~\exp\left(i\sum_{i} \tilde\rho_{i\dot\alpha} \tilde\mu_i^{\dot\alpha}\right)\exp\left(i\sum_{i} \rho_{i\alpha}\mu_i^\alpha \right) \\
&~~~~~~~~~~~~~~~~~~\delta^{2,n}(\tilde \rho_{i\dot\alpha} - c_{ji}\tilde \lambda_{j\dot\alpha} ) \delta^{2,n}(\rho_{i\alpha} - c_{ij} \lambda_{j\alpha} ) \mathcal{F}_n^{h_1,\cdots, h_n}(c_{mn}).
\end{align}
The $\rho \;\textrm{and}\; \tilde \rho$ integrals can be performed using the delta functions, and we obtain
\begin{align}
    \notag\hat{\psi}^{h_1\cdots h_n}_n(Z,W) =& \int dc_{mn} \exp\left( i\sum_{i,j} c_{ij}(\lambda_i \cdot \mu_j + \tilde\lambda_j \cdot \mu_i)  \right) \mathcal{F}^{h_1\cdots h_n}_n(c_{mn})\\
    =& \int dc_{mn} \exp\left( i\sum_{i,j} c_{ij} Z_i\cdot W_j  \right) \mathcal{F}^{h_1\cdots h_n}_n(c_{mn})
\end{align}
Hence, the ambitwistor correlator can be simply  expressed as
\begin{align}\label{AmbiNPoint}
\boxed{\hat{\psi}^{h_1\cdots h_n}_n(Z, W) =\int d^{n^2}c_{mn} \,
  e^{ic_{ij}Z_{i} \cdot W_j}\mathcal{F}^{h_1\cdots h_n}_n(c_{mn}).}
\end{align}
which matches the general ansatz we started with in the previous section, but the function \(\mathcal{G}^{h_1\cdots h_n}_{s_1\cdots s_n}\) is now dictated by the \(\mathcal{F}^{h_1\cdots h_n}_n(c_{mn})\) of the Grassmannian answers. The expressions for the correlators obtained by this method match those derived from the conformal bootstrap. This agreement provides a nontrivial consistency check of the formalism and confirms that the twistor-space construction correctly translates to the known momentum-space correlators.

With the connection to momentum-space established, we now take a minor detour to discuss the parity-odd structures in twistor space. 

\subsection{Parity-Odd Correlators in Ambitwistor space}\label{sec:parity-odd}
Recall that in eq \eqref{AmbiSH}, the momentum vector is written in terms of the spinor-helicity variables as  
    \begin{equation}\label{pmatrix}
        p_{\alpha\dot\alpha} = \lambda_{\alpha} \tilde\rho_{\dot\alpha} - \rho_{\alpha}\tilde\lambda_{\dot\alpha} = \left(
\begin{array}{cc}
 -p_t-p_z & p_x-p_y \\
 p_x+p_y & p_t-p_z \\
\end{array}
\right) .
    \end{equation}
In the \((2,2)\) signature, parity acts as \( p_y \to -p_y\), which, in the above choice of bispinor conventions, is realized by simply \textit{transposing} the matrix.
It is, therefore, natural to write the parity-transformed momentum as 
\begin{equation}
\mathcal{P}(p_{\alpha\dot\alpha}) = p_{\dot\alpha\alpha} = \lambda_{\dot\alpha} \tilde\rho_{\alpha} - \rho_{\dot\alpha}\tilde\lambda_{\alpha}.
\end{equation}
Comparing with the original momentum bispinor, eq \eqref{pmatrix} one can infer that the parity transformation is equivalent to the following exchanges 
\begin{equation}\label{parity-exchanges}
    \mathrm{Parity}:\qquad \lambda_{\alpha} \to \tilde\lambda_{\alpha}, ~\tilde\lambda_{\dot\alpha}\to -\lambda_{\dot\alpha},~\rho_{\alpha}\to \tilde\rho_{\alpha}, ~\tilde\rho_{\dot\alpha} \to -\rho_{\dot\alpha}.
\end{equation}
Doing a half-Fourier transform, under parity we have 
\begin{equation}
    \textrm{Parity:}\qquad \mathcal{P}(Z) = \{\tilde\lambda_{\alpha},-\mu^{\dot\alpha}  \},\qquad \mathcal{P}(W) = \{\tilde\mu^{\alpha}, -\lambda_{\dot\alpha}  \},
\end{equation}
and therefore, the invariant dot product becomes 
\begin{equation}
    \mathcal{P}(Z_i\cdot W_j) = \tilde\lambda_i \cdot \tilde\mu_j + \lambda_j \cdot \mu_i \equiv  Z_j\cdot W_i.
\end{equation}
Given the general ansatz eq \eqref{ansatz}, this transformation can alternatively be realized by applying the variable change \(c_{ij}\leftrightarrow c_{ji}\). Therefore, the action of parity on the function in Schwinger parameter space is to simply exchange all the \(c_{ij}\) with \(c_{ji}\)
\begin{align}\label{cijparity}
   \mathcal{P}(c_{ij})=c_{ji},\qquad\forall\;i,j.
\end{align}

In twistor space, we observe the following behavior of correlators in eq \eqref{ansatz} under the above exchange eq \eqref{cijparity}: for parity-even correlators, we have \footnote{The action of parity as given in eq \eqref{parity-exchanges} is one of the multiple equivalent choices of exchanges, with this particular choice being especially convenient since the parity becomes a spin-dependent statement here. In another choice, \(\lambda\leftrightarrow\tilde\lambda,~\rho\leftrightarrow - \tilde\rho\), the parity in Schwinger parameters would look like \(c_{ij}\leftrightarrow - c_{ji}\), and in this case the action of parity would be spin-independent, but would become a helicity-dependent statement.}
\begin{equation}
    \mathcal{P}\big(\braket{O_{s_1}O_{s_2}O_{s_3}}_e\big) = (-1)^{s_1 + s_2 + s_3}  \braket{O_{s_1}O_{s_2}O_{s_3}}_e,
\end{equation}
and for parity-odd correlators, we obtain
\begin{equation}
    \mathcal{P}\big(\braket{O_{s_1}O_{s_2}O_{s_3}}_o\big) = -(-1)^{s_1 + s_2 + s_3}  \braket{O_{s_1}O_{s_2}O_{s_3}}_o.
\end{equation}
This is an empirical observation one can infer from the correlators that have only parity-even structures in table \ref{tab:correlators}, and is the same as what \cite{CarrilloGonzalez:2026eum} observe.\\

With the above action of parity, we can immediately see that the structures in table \ref{tab:correlators} for \(\braket{JJJ}, ~\braket{TJJ}\) and \(\braket{TTT}\), accompanied by the coefficients \(\mathcal{C}_i\), are parity-even structures, while those accompanied by \(\mathcal{D}_i\) are parity-odd.\\


\noindent\(\mathbf{\braket{JJJ}_{Odd}}\)\\
Let us consider the parity-odd part of the \(\braket{J^+J^+J^+}\) correlator in table \ref{tab:correlators}. This is given by 
\begin{align}\label{JJJOdd}
    \mathcal{G}_{JJJ} = \frac{(c_{12} c_{23} c_{31} + c_{13} c_{32} c_{21})}{(c_{12} c_{23} c_{31} - c_{13} c_{32} c_{21})^3}\Sgn(c_{12} c_{23} c_{31} + c_{13} c_{32} c_{21}).
\end{align}
Under exchange of two legs, we observe that this correlator \textit{does not} pick a minus sign.

Parity-odd correlators in momentum space \cite{Bzowski:2013sza,Coriano:2013jba,Bzowski:2017poo,Bzowski:2018fql} take simple form \cite{Jain:2021wyn,Jain:2021vrv,Jain:2021gwa,Coriano:2024ssu,Coriano:2026zag}. Consider the momentum-space expression corresponding to the parity-odd structure eq \eqref{JJJOdd}, which can be obtained by using eq \eqref{JJJOdd} in the Grassmannian ansatz eq \eqref{eq:grassmannian-correlator}, and performing the \(c_{ij}\) integrals as outlined in \cite{Bala:2026trw}.  The resultant momentum-space correlator that we obtain is
\begin{align}
    \notag\braket{J_\mu^a J_\nu^b J_\rho^c} = d^{abc}\frac{k_1^2 k_2^2 k_3^2}{(J_2)^{5/2}}\Big(
\epsilon(\epsilon_1,\epsilon_2,\epsilon_3,k_1)
(k_1^2-k_2^2-k_3^2)(k_2^2-k_3^2)
+\text{cyclic}
\Big),
\end{align}
where \({J}_2 = -\left(k_1+k_2+k_3\right) \left(k_1+k_2-k_3\right) \left(k_1-k_2+k_3\right) \left(-k_1+k_2+k_3\right)\) and \(d^{abc}\) is the totally symmetric invariant tensor of the gauge group. The kinematic factor in this expression is symmetric under permutations of the three external legs, in agreement with the exchange symmetry of the corresponding twistor-space structure. More importantly, the resulting expression has a direct relation to the momentum-space correlator obtained from a bulk computation. In particular, it precisely agrees with the discontinuity in \(k_1\) and \(k_3\) of the parity-odd \(\braket{JJJ}\) correlator obtained from doing a bulk Witten diagram computation of Chern-Simons theory in AdS\(_5\) bulk\footnote{See appendix \ref{CST} for details of the bulk computation.}. The same expression can alternatively be obtained from a free chiral Fermion theory in four dimensions. Thus, the momentum-space provides an independent check of the parity assignment. \\

\noindent{$\mathbf{\braket{TJJ}_{Odd}}$}\\
If we now consider the parity-odd part $\mathcal{G}(c_{mn})$ of the $\braket{TJJ}$ correlator, which is given in table \ref{tab:correlators} to be 
\begin{equation}
    \mathcal{G}_{TJJ} = \frac{c_{23} c_{32} \left(c_{12} c_{23} c_{31}+c_{13} c_{21} c_{32}\right) }{\left(c_{12} c_{23} c_{31}-c_{13} c_{21} c_{32}\right){}^4}\Sgn(c_{12} c_{23} c_{31}-c_{13} c_{21} c_{32}),
\end{equation}
we find that it is antisymmetric under the exchange $2\leftrightarrow 3$, while \(\braket{TJJ}\) should be symmetric under the exchange \cite{Jain:2021wyn}. Thus, although conformal symmetry and current conservation admit a parity-odd tensor structure for $\braket{TJJ}$, it is eliminated upon imposing the required exchange symmetry of the identical currents $J$, and therefore, the physical parity-odd $\braket{TJJ}$ correlator vanishes\footnote{The counting \(2\min(s_1,s_2,s_3)+1\) discussed in section \ref{Correlators} follows from conformal symmetry and current conservation alone. When two or more external operators are identical, their required exchange symmetry imposes additional constraints, reducing the number of physically allowed structures.}.\\

\noindent{$\mathbf{\braket{TTT}_{Odd}}$}\\
A similar argument applies to $\braket{TTT}$ parity-odd, which is given by 
\begin{align}
    \notag\mathcal{G}_{TTT} &=  \mathcal{D}_1 \frac{(c_{12} c_{23} c_{31}+c_{13} c_{21} c_{32})}{ \left(c_{12} c_{23} c_{31}-c_{13} c_{21} c_{32}\right){}^4}\Sgn(c_{12} c_{23} c_{31}-c_{13} c_{21} c_{32})  \\
    &~~~~~+ \mathcal{D}_2 \frac{c_{12} c_{13} c_{21} c_{23} c_{31} c_{32} \left(c_{12} c_{23} c_{31}+c_{13} c_{21} c_{32}\right)}{ \left(c_{12} c_{23} c_{31}-c_{13} c_{21} c_{32}\right){}^6}\Sgn(c_{12} c_{23} c_{31}-c_{13} c_{21} c_{32}).
\end{align}
Although conformal symmetry and conservation allow parity-odd tensor structures, the required Bose symmetry under exchange of the identical stress tensors eliminates those that are antisymmetric, so the physical parity-odd $\braket{TTT}$ correlator vanishes \cite{Jain:2021wyn}. But if one were to compute the correlator for some conserved spin-2 current carrying some index, which would allow for the antisymmetry under the exchange, then such a correlator will have physical parity-odd structures, which would be expected to match the structure obtained from the above \(\mathcal{G}_{TTT}\)

Until now, we have developed the twistor-space representations of conformal correlators, formalizing their generators and their corresponding correlation functions. We now present the method to obtain the familiar position-space correlators, starting from their twistor analogues. We do so by developing the Penrose transform, which will be the subject of our discussion in the next section.

\subsection{Penrose Transform for Ambitwistors}\label{AmbiPenrose}
The Penrose transform is an integro-differential transformation that maps twistor space expressions to their position-space representatives, subject to certain constraints. In this section, we present the derivation of the Penrose transform for ambitwistors and then obtain position-space correlators starting from twistor-space correlators.

\subsubsection{Derivation of Penrose Transform}
We closely follow the method outlined for CFT\(_3\) in \cite{Bala:2025qxr} to derive the Penrose transform for a conserved spin-1 current in CFT\(_4\). The same method can then be generalised to other spinning currents too.
         The three dimensional space orthogonal to a momentum vector \(p_{\alpha\dot\alpha}\) is spanned by the following basis eq \eqref{eq:Helicity-basis-spinors}
         \begin{equation}
             \xi^+_{\alpha \dot\alpha} = \frac{\lambda_{\alpha} \tilde\lambda_{\dot \alpha}}{p},~~~ \xi^0_{\alpha \dot\alpha} = \frac{\lambda_{\alpha}\tilde\rho_{\dot\alpha} + \rho_{\alpha}\tilde\lambda_{\dot\alpha}}{p},~~~ \xi^-_{\alpha \dot\alpha} = \frac{\rho_{\alpha}\tilde\rho_{\dot\alpha}}{p}.
         \end{equation}
         
        Starting with the definition of the Fourier transform, we have
        \begin{align}
            \notag J_{\alpha\dot\alpha}(x) &
            = \int \frac{d^2\rho\, d^2\tilde\rho\, d^2\lambda \, d^2\tilde\lambda}{\text{Vol}(SL(2,\mathbb{R}))\text{Vol}(GL(1,\mathbb{R}))} e^{i p\cdot x} \epsilon_{\alpha\beta}\epsilon_{\dot\alpha\dot\beta} J^{\beta\dot\beta}(p)\\
             &= \int \frac{d^2\rho\, d^2\tilde\rho\, d^2\lambda \, d^2\tilde\lambda}{\text{Vol}(SL(2,\mathbb{R}))\text{Vol}(GL(1,\mathbb{R}))} e^{i p\cdot x} \frac{p_{\alpha\dot\alpha}p_{\beta \dot\beta} - p_{\alpha\dot\beta}p_{\beta\dot\alpha}}{p^2} J^{\beta\dot\beta}(p),
        \end{align}
        where we have used the identity \(p_{\alpha\dot\alpha}p_{\beta \dot\beta} - p_{\alpha\dot\beta}p_{\beta\dot\alpha} = p^2 \,\epsilon_{\alpha\beta}\epsilon_{\dot\alpha\dot\beta}  \).
        This is, in essence, a decomposition of the vector \(J_{\alpha\dot\alpha}\) into the components along and orthogonal to \(p_{\alpha\dot\alpha}\). The first term, which projects to the component along \(p_{\alpha\dot\alpha}\), drops off due to the conservation equation \(p_{\beta\dot\beta}J^{\beta\dot\beta} = 0\), while the second term can be expanded and written in terms of the polarization basis\footnote{The rescaled conserved current is defined as $\hat{J}_s=\frac{J_s}{p^s}$. This rescaling ensures a simple action of the SCT generator for CFT$_4$, much like in CFT$_3$ \cite{Maldacena:2011nz}, where it was rescaled with $\frac{1}{p^{s-1}}$. However, it is interesting to note that the correct rescaling for 4-d is naturally obtained while deriving the Penrose transform.} as
        \begin{align}
            \notag J_{\alpha\dot\alpha}(x) 
             &=\int \frac{d^2\rho\, d^2\tilde\rho\, d^2\lambda \, d^2\tilde\lambda}{\text{Vol}(SL(2,\mathbb{R}))\text{Vol}(GL(1,\mathbb{R}))} e^{i p\cdot x} \left( \rho_{\alpha}\tilde\rho_{\dot\alpha}\, \hat J^+(p)  + \lambda_{\alpha}\tilde\lambda_{\dot\alpha} \, \hat J^-(p) - (\lambda_{\alpha} \tilde\rho_{\dot\alpha} + \rho_{\alpha}\tilde\lambda_{\dot\alpha})\hat J^0(p)  \right) .
        \end{align}
        Having isolated the three polarization components, we now express the corresponding components as a half-Fourier transform from twistor space separately
        \begin{align}
            \notag J_{\alpha\dot\alpha}(x)& = \int d^2\mu\,d^2\tilde\mu\, \frac{d^2\rho\, d^2\tilde\rho\, d^2\lambda \, d^2\tilde\lambda}{\text{Vol}(SL(2,\mathbb{R}))\text{Vol}(GL(1,\mathbb{R}))}  \big( \rho_{\alpha}\tilde\rho_{\dot\alpha}\, \hat J^+(Z,W)  + \lambda_{\alpha}\tilde\lambda_{\dot\alpha} \, \hat J^-(Z,W) \\ &\hspace{120pt}- (\lambda_{\alpha} \tilde\rho_{\dot\alpha} + \rho_{\alpha}\tilde\lambda_{\dot\alpha})\hat J^0(Z,W)  \big) e^{-i\rho\cdot \mu - i\tilde\rho \cdot \tilde\mu } e^{i \lambda \cdot x \cdot \tilde\rho- i\rho\cdot x \cdot \tilde\lambda}\label{eq:Ambi-twistor-ans-hft}\\
            \notag &= \int d^2\mu\,d^2\tilde\mu\, \frac{d^2\rho\, d^2\tilde\rho\, d^2\lambda \, d^2\tilde\lambda}{\text{Vol}(SL(2,\mathbb{R}))\text{Vol}(GL(1,\mathbb{R}))}  \Big( -\hat J^+(Z,W)\frac{\partial}{\partial \mu^\alpha}\frac{\partial}{\partial \tilde\mu^{\dot\alpha}}  + \lambda_{\alpha}\tilde\lambda_{\dot\alpha} \, \hat J^-(Z,W) \\& \hspace{120pt}-i\,\hat J^0(Z,W) \Big(\lambda_{\alpha} \frac{\partial}{\partial \tilde\mu^{\dot\alpha}}  +\tilde\lambda_{\dot\alpha}\frac{\partial}{\partial \mu^{\alpha}} \Big)  \big) e^{-i\rho\cdot \mu - i\tilde\rho \cdot \tilde\mu } e^{i \lambda \cdot x \cdot \tilde\rho- i\rho\cdot x \cdot \tilde\lambda}.
        \end{align}
        Performing integration by parts transfers the \(\mu\) and \(\tilde\mu\) derivatives onto  \(J^{\pm,0}\). The subsequent \(\rho\) and \(\tilde\rho\) integrals simply generate the incidence-relation delta functions
        \begin{equation}
           \delta^2\left(\mu^{\alpha}+x^{\alpha\dot\alpha} \tilde\lambda_{\dot\alpha}\right)~~~~\&~~~~ \delta^2\left(\tilde\mu^{\dot\alpha} - \lambda_{\alpha}x^{\alpha\dot\alpha} \right).
        \end{equation}
        Thus, the Penrose transform for the conserved spin-1 current finally takes the form
       \begin{align}\label{eq:Ambi-penrose-spinning}
    \boxed{
    \begin{aligned}
    J_{\alpha\dot\alpha}(x)
    &=
    \int \frac{d^2\lambda \, d^2\tilde\lambda}
    {\mathrm{Vol}(GL(1,\mathbb{R}))^2}
    \Bigg(
    -\frac{\partial}{\partial \mu^\alpha}
    \frac{\partial}{\partial \tilde\mu^{\dot\alpha}}
    \hat J^+(Z,W)
    +\lambda_{\alpha}\tilde\lambda_{\dot\alpha}\hat J^-(Z,W)
    \\
    &\hspace{120pt}
    +i\,\Big(
    \lambda_{\alpha}\frac{\partial}{\partial\tilde\mu^{\dot\alpha}}
    +\tilde\lambda_{\dot\alpha}\frac{\partial}{\partial\mu^\alpha}
    \Big)\hat J^0(Z,W)
    \Bigg)
    \Bigg|_{\substack{
    \tilde\mu^{\dot\alpha}=\lambda_\alpha x^{\alpha\dot\alpha}\\
    ~~\mu^\alpha=-x^{\alpha\dot\alpha}\tilde\lambda_{\dot\alpha}
    }} .
    \end{aligned}
    }
    \end{align}
A few comments are in order. Notice that in eq \eqref{eq:Ambi-penrose-spinning}, we divide with \(\textrm{Vol}(GL(1,\mathbb{R}))^2\), instead of \(\textrm{Vol}(SL(2,\mathbb{R})) \Vol(GL(1,\mathbb{R}))\). This is because by selectively performing the \(\rho\) and \(\tilde\rho\) integrals, the \(SL(2,\mathbb{R})\) invariance is manifestly lost, except for the diagonal part of it. However, the \(GL(1,\mathbb{R})\) invariance part remains intact\footnote{As an illustrative example, consider a  \(GL(1,\mathbb{R})\times SL(2,\mathbb{R})\) invariant object, \(\exp(i p^2) \equiv \exp(i(\lambda\cdot \rho)(\tilde\lambda\cdot \tilde\rho))\). Doing its half-Fourier transform gives, 
\begin{align}
    \notag \int d^2\rho~d^2\tilde\rho \, e^{(i\mu\cdot \rho + i\tilde\mu\cdot \tilde \rho)} e^{i(\lambda\cdot \rho)(\tilde\lambda\cdot \tilde\rho)} = \delta\left(\mu^1 \lambda_2 -\mu^2 \lambda_1\right)\delta\left(\tilde\mu^{1}\tilde\lambda_{2} -\tilde\mu^{2} \tilde \lambda_{1}\right)\exp\left(- i \frac{\mu^1 \tilde\mu^{2}}{\lambda_1 \tilde\lambda_{2} }  \right).
\end{align}
The above expression can be found not to be annihilated by the off-diagonal part of \(SL(2,\mathbb{R})\), whereas the diagonal \(SL(2,\mathbb{R})\) and \(GL(1,\mathbb{R})\), which are rescaling equations, annihilate it. }.
    
Moreover, the above construction extends directly to conserved currents of arbitrary integer spin. With the Penrose transform eq \eqref{eq:Ambi-penrose-spinning} for the ambitwistor setup with us, let us now compute the position-space correlators from their twistor-space counterparts.

\subsubsection{Position-Space Results}\label{sec:Pen}
Using the Penrose transform eq \eqref{eq:Ambi-penrose-spinning} derived in the previous section, one can proceed to convert the ambitwistor correlators to position-space. We now present the results for a few two- and three-point functions. 
\subsubsection*{Two-Point Functions}  
\noindent\(\mathbf{\braket{O_2O_2}}\)\\
    Let us start with the scalar two-point function for operator $O_2$. Starting with the ansatz for \(\braket{O_2O_2}\) given in eq \eqref{eq:o2o2-ambitwistor}, and using the Penrose transform eq \eqref{eq:Ambi-penrose-spinning}, we have 
    \begin{equation}
        \braket{O_2(x_1) O_2(x_2)} = \int \frac{d^2\lambda_1\, d^2\lambda_2\, d^2\tilde\lambda_1\, d^2\tilde\lambda_2}{\text{Vol}(GL(1,\mathbb{R}))^4}\, \delta(Z_1\cdot W_1) \, \delta(Z_2\cdot W_2) \, \delta(Z_1\cdot W_2)\, \delta(Z_2\cdot W_1) \Bigg|_{X}.
    \end{equation}
    When subjected to the incidence relations, the twistor dot product \(Z_i\cdot W_j\) becomes 
    \begin{align}
        Z_i \cdot W_j \Big|_{X} = - \lambda_{i\alpha} x_j^{\alpha\dot\alpha} \tilde\lambda_{j\dot\alpha} + \lambda_{i\alpha}x_i^{\alpha\dot\alpha} \tilde\lambda_{j\dot\alpha}
        = \lambda_{i\alpha}x_{ij}^{\alpha\dot\alpha}\tilde\lambda_{j \dot\alpha}.
    \end{align}
    Therefore, the scalar two-point in position-space becomes \footnote{
    Upon imposing the incidence relations, \(\delta(Z_1\cdot W_1)\) and \(\delta(Z_2\cdot W_2)\) both go to \(\delta(0) = \int dc \exp(0) = \mathrm{Vol}(GL(1, \mathbb{R}))\). Therefore, they cancel two volume factors.
    }
    \begin{align}
        \notag\braket{O_2(x_1) O_2(x_2)} &= \int \frac{d^2\lambda_1\, d^2\lambda_2\, d^2\tilde\lambda_1\, d^2\tilde\lambda_2}{\text{Vol}(GL(1,\mathbb{R}))^2}\, \\
       &~~~~~~~~~~~ \int dc_{12}dc_{21}\, \exp \left(ic_{12}\lambda_1\cdot x_{12}\cdot \tilde\lambda_2\right)\, \exp\left(ic_{21}\lambda_2\cdot x_{21}\cdot\tilde\lambda_1\right),
    \end{align}
    where we have now written the delta functions as Schwinger integrals.
    
    We can now do the \(\lambda\) and \(\tilde\lambda\) integrals, which leads to the following expression
    \begin{equation}
        \braket{O_2(x_1) O_2(x_2)} =\frac{1}{\textrm{Det}(x_{12})} \frac{1}{\textrm{Det}(x_{21})} \frac{1}{\text{Vol}(GL(1,\mathbb{R}))^2}\, \int dc_{12}dc_{21}\, \frac{1}{c_{12}^2c_{21}^2}. 
    \end{equation}
    By a change of variable \(c \to 1/u\), we can show that 
    \begin{equation}
        \int_{-\infty}^{\infty} \frac{dc}{c^2} \to -\int_{0^-}^{-\infty} du \,-\int_{\infty}^{0^+} du = \int_{-\infty}^{\infty}du = \Vol(GL(1,\mathbb{R}))\label{eq:vol-gl1r},
    \end{equation}
    and therefore, the remaining two integrals produce two Vol\((GL(1,\mathbb{R}))\) factors, completely canceling the volume factors in the denominator. Therefore, the Penrose transform results in the correct position-space expression for the scalar two-point function 
    \begin{equation}
        {\braket{O_2(x_1) O_2(x_2)} = \frac{1}{x_{12}^4}.}
    \end{equation}\\

\noindent\(\mathbf{\braket{JJ}}\)\\    
    Here we present a general method to perform the Penrose transform for spinning correlators, with \(\braket{JJ}\) as an illustrative example. The same method will be generalised to the case of three-point correlators too. 
    For spinning correlators, it is convenient to work directly with the Schwinger parameter representation of the correlators. Under the Penrose transform eq \eqref{eq:Ambi-penrose-spinning}, factors of $\lambda$ and $\tilde\lambda$ appear either explicitly or through derivatives with respect to $\mu$ and $\tilde\mu$. After imposing the incidence relations, the resulting integrals take the generic form
    \begin{equation}
        \int d^2\lambda_i d^2\tilde\lambda_i \, dc_{ij}\, P(c_{ij}, \lambda_i,\tilde\lambda_j) \exp\left( i c_{ij}\,\lambda_i \cdot x_{ij} \cdot \tilde\lambda_j  \right), 
    \end{equation}
    where \(P\) is some polynomial in \(\lambda\) and \(\tilde \lambda\). Such integrals are most conveniently evaluated by introducing source terms for \(\lambda_{i\alpha}\tilde\lambda_{j\dot\alpha}\), doing the spinor integrals, and then performing the required differentiations with respect to the source. For the correlator \(\braket{J^+ J^+}\), we have
    \begin{equation}
    \braket{J_{\alpha\dot\alpha}^+(Z_1, W_1) J_{\beta\dot\beta}^+(Z_2, W_2)} = \int dc_{ij} \exp(i\,c_{ij} \, Z_i\cdot W_j) \frac{1}{c_{12}c_{21}}.
    \end{equation}
    Under the Penrose transform eq \eqref{eq:Ambi-penrose-spinning}, we get 
    \begin{align}
        \notag \braket{J_{\alpha\dot\alpha}^+(x_1) J_{\beta\dot\beta}^+(x_2)} = \int\, \frac{d^2\lambda_i d^2\tilde\lambda_i}{\mathrm{Vol}(GL(1, \mathbb R))^4}\, dc_{ij}\, \left(c_{11} \lambda _{1\alpha}+c_{21} \lambda _{2\alpha}\right) \left(c_{11} \tilde\lambda_{1\dot\alpha}+c_{12} \tilde\lambda_{2\dot\alpha}\right)~~~~~~~~
        \\\left(c_{12} \lambda _{1\beta}+c_{22} \lambda _{2\beta}\right) \left(c_{21} \tilde\lambda_{1\dot\beta}+c_{22} \tilde\lambda_{2\dot\beta}\right)
        \exp\left(i \,c_{ij}\,Z_i\cdot W_j \right)\frac{1}{c_{12}c_{21}}\Bigg|_{X}.
\label{eq:JJpp-Penrose}
    \end{align}
    Before proceeding, notice that any term in the above equation that is of the form 
    \begin{equation}
        \int\prod_i d^2  \lambda_i \, d^2\tilde\lambda_i \prod_{ij}\,dc_{ij} \, c_{kk}^m \exp(\sum_{ij}ic_{ij}Z_i\cdot W_j) f(c_{ij}, \lambda_i, \tilde\lambda_i)\Bigg|_{X},
    \end{equation}
    can be written as 
    \begin{equation}
        \int d^2 \lambda_i d^2\tilde\lambda_idc_{ij}\left(\left( \frac{-i\eta_a}{\eta\cdot \mu_k}\frac{\partial}{\partial \lambda_{ka}}\right)^m \exp(ic_{kk}Z_k\cdot W_k)\right) \exp\left(i\sum_{\substack{i\ne k \\ j\ne k}}c_{ij}Z_i\cdot W_j\right) f(c_{ij}, \lambda_i, \tilde\lambda_i)\Bigg|_{X}.
    \end{equation}
    At the support of the incidence relations, \(Z_k\cdot W_k = 0\), and therefore, these terms will involve taking derivatives of a constant and therefore evaluate to zero. So we can drop all the diagonal terms from the Penrose transform. This ties back to our statement in section \ref{Correlators}, that the \(\delta(Z_i\cdot W_i)\) and their derivatives do not contribute to the tensor structures, and therefore act as trivial constraints under the Penrose transform.\\
    
    Now let us return to the integrand in hand eq \eqref{eq:JJpp-Penrose}. Dropping the diagonal terms, thereby canceling two volume factors, we get 
    \begin{align}
        \notag &\braket{J_{\alpha\dot\alpha}^+(x_1) J_{\beta\dot\beta}^+(x_2)} =\\&~~~~~~ \int\, \frac{d^2\lambda_i d^2\tilde\lambda_i}{\mathrm{Vol}(GL(1, \mathbb R))^2}\, dc_{12}dc_{21}\, \lambda _{2\alpha}\tilde\lambda_{2\dot\alpha} \lambda _{1\beta}\tilde\lambda_{1\dot\beta}  \exp\left(i \,c_{12}\,\lambda_1\cdot x_{12} \cdot \tilde\lambda_2  + i c_{21}\,\lambda_2\cdot x_{21} \cdot \tilde\lambda_1 \right)c_{12}c_{21}.
    \end{align}
The integrand possesses scaling redundancy where the spinors \(\lambda_i\)(\(\tilde\lambda_i\)) and the coefficients \(c_{ij}\) can be rescaled simultaneously without changing the integrand. To isolate the non-redundant degrees of freedom, it is convenient to perform a change of variables in which the coefficients in the exponent are absorbed into the spinors, thereby making the scaling directions explicit. With this change of variables, the correlator becomes
    \begin{align}
        \notag \braket{J_{\alpha\dot\alpha}^+(x_1) J_{\beta\dot\beta}^+(x_2)} = \int \frac{dc_{12}}{c_{12}^2} \frac{dc_{21}}{c_{21}^2}\int\, \frac{d^2\lambda_i d^2\tilde\lambda_i}{\mathrm{Vol}(GL(1, \mathbb R))^2}\,  \lambda _{2\alpha}\tilde\lambda_{2\dot\alpha} \lambda _{1\beta} \tilde\lambda_{1\dot\beta}\\
        \exp\left(i \,\lambda_1\cdot x_{12} \cdot \tilde\lambda_{2} + i\, \lambda_2\cdot x_{21} \cdot \tilde\lambda_1 \right).
    \end{align}
    We can see immediately that the two volume factors in the denominator cancel the two volume factors from the \(c_{12}\) and \(c_{21}\) integrals, which are the same as in eq \eqref{eq:vol-gl1r}, giving us a finite tensor structure. 
    To evaluate the spinor integrals, we now introduce source terms \(i\,\lambda_{i\alpha} S_{ij}^{\alpha\dot\alpha} \tilde\lambda_{j\dot\alpha}\), and define the generating function
    \begin{equation}
        Z[S] = \int d\lambda_i d\tilde\lambda_i   \exp\left(i~\lambda_{i\alpha} (x_{ij} + S_{ij})^{\alpha\dot\alpha} \tilde\lambda_{j\dot\alpha} \right) = \frac{1}{\det(X + S)}.
    \end{equation}
    In terms of the generating function, the correlator can be written as   
    \begin{align}          
    \notag\braket{J^+(x_1)J^+(x_2)} &= (-i)^2\partial_{S_{22}^{\alpha\dot\alpha}}\partial_{S_{11}^{\beta\dot\beta}}Z[S]\Big|_{S=0}. 
    \end{align}    
    Evaluating the derivatives\footnote{
        We use the following standard identities 
        \begin{equation}
            \frac{d}{dS_{ij}^{a\dot a}} \frac{1}{\det (X+S)} = -\frac{1}{\det (X+S)} (X+S)^{-1}_{j\dot a, i a}~~~~\&~~~~
            \frac{d}{dS_{kl}^{b\dot b}} (X+S)^{-1}_{j\dot a, ia} = - (X+S)^{-1}_{j\dot a, k b} (X+S)^{-1}_{l\dot b, i a}
        \end{equation}
        which are true when \(X\) is independent of \(S\), using which one can write for \(n\) derivatives acting on \(Z[S]\), 
        \begin{equation}\label{eq:source-n-derivative}
            \frac{d^n}{ dS_{i_1 j_1}^{a_1 \dot a_1}\cdots dS_{i_n j_n}^{a_n \dot a_n}} \frac{1}{\det(X+S)}\bigg|_{S=0} = (-1)^n \frac{1}{\det(X)}\sum_{\sigma\in \mathrm{Sym}_n} X^{-1}_{j_1 \dot a_1, i_{\sigma(1)} a_{\sigma(1)} } X^{-1}_{j_1 \dot a_1, i_{\sigma(2)} a_{\sigma(2)} } \cdots X^{-1}_{j_1 \dot a_1, i_{\sigma(n)} a_{\sigma(n)} }
        \end{equation}
        The permutations will inevitably result in tensor structures with mixed indices of the form \(x_{ij\, a\dot b} x_{kl\, b\dot a} \) etc, which should then be converted to \textit{un-mixed} form by using the Pauli matrix identities of the following form 
        \begin{equation}
            \sigma^\mu_{a\dot a} \sigma^\nu_{b\dot b} + \sigma^\mu_{b\dot b} \sigma^\nu_{a\dot a} = 2\eta^{\mu\nu} + \sigma^\mu_{a\dot b} \sigma^\nu_{b\dot a} +\sigma^\mu_{b\dot a} \sigma^\nu_{a\dot b}.
        \end{equation}
    }, one obtains the correct position-space correlator
    \begin{equation}\label{J+J+}
        {\braket{J_{\alpha\dot\alpha}^+(x_1)J_{\beta\dot\beta}^+(x_2)} =  \frac{1}{x_{12}^6}\left(\epsilon_{\alpha\beta}\epsilon_{\dot\alpha\dot\beta}-\frac{x_{12 \alpha \dot\alpha} x_{12 \beta \dot\beta}}{x_{12}^2}\right) = \frac{\sigma_{\alpha\dot\alpha}^{\mu} \sigma_{\beta\dot\beta}^{\nu}}{x_{12}^6}\left( \frac{\eta_{\mu\nu}}{2} - \frac{x_{12\mu}x_{12\nu}}{x_{12}^2}   \right).} 
    \end{equation}
    One can instead seek to obtain the position-space correlator via the Penrose transform of the \(\braket{J^- J^-}\) correlator. The ambitwistor ansatz for the \(\braket{J^-J^-}\) correlator is 
    \begin{equation}
        \int dc_{ij} \exp(ic_{ij}Z_i\cdot W_j) \frac{\left(c_{12} c_{21}-c_{11} c_{22}\right){}^2}{2 c_{12} c_{21}}.
    \end{equation}
    Dropping the diagonal terms and putting this inside the Penrose transform, we see that the expression reduces exactly to that of the \(\braket{J^+ J^+}\).  Thus, only a single helicity component is enough to reproduce the entire position-space correlator. The same trend of producing the full position-space answer from only a single helicity component (as was the case in CFT$_3$ \cite{Bala:2025qxr}) has been observed in correlators of higher spins and at higher point too.

    We now extend our discussion to three-point correlators.
\subsubsection*{Three-Point Functions} 
\(\mathbf{\braket{O_2O_2O_2}}\)\\
    Let us begin with the scalar three-point function. The twistor-space ansatz eq \eqref{eq:Ambi-o2o2o2}, when expressed as a Schwinger integral, is of the following form
    \begin{equation}
        \braket{O_2(Z_1, W_1)O_2(Z_2, W_2)O_2(Z_3, W_3)} = \int dc_{ij} \, \exp\left( i\, c_{ij}\, Z_i\cdot W_j \right) \frac{1}{|c_{12}c_{23}c_{31} - c_{13}c_{32}c_{21}|}. 
    \end{equation}
    The Penrose transform eq \eqref{eq:Ambi-penrose-spinning} is simply 
    \begin{equation}
        \braket{O_2(x_1)O_2(x_2)O_2(x_3)} = \int \frac{d\lambda_id\tilde\lambda_i}{\mathrm{Vol}(Gl(1,\mathbb{R}))^3}dc_{ij}\, \exp\left( ic_{ij} \lambda_i\cdot x_{ij}\cdot \tilde\lambda_j \right) \frac{1}{|c_{12}c_{23}c_{31} - c_{13}c_{32}c_{21}|}. 
    \end{equation}
   It is straightforward to perform the \(\lambda\) and \(\tilde\lambda\) integrals, which results in 
    \begin{equation}
      {\braket{O_2(x_1)O_2(x_2)O_2(x_3)} = \mathcal{N}_{OOO}\frac{1}{x_{12}^2 x_{23}^2 x_{31}^2} ,}
    \end{equation}
    where \(\mathcal{N}_{OOO}\) is \(x\)-independent normalization\footnote{The three-point functions seem to have more volume factors arising from the spinor integrals than there are in the denominator to cancel, and the origin of these extra divergences is unclear at this moment.}.\\
    
\noindent \(\mathbf{\braket{JJO_2}}\)\\
    We now move on to the correlators involving spinning operators. Spinning correlators that admit a unique conformally invariant conserved tensor structure can be treated in complete analogy with spinning two-point correlators. In these cases, the dependence on the \(c_{ij}\) parameters factorizes from the tensor structure and can therefore be absorbed into the overall normalization. To illustrate, let us look at the case of the correlator \(\braket{JJO}\). \\
    The ambitwistor-space ansatz for this correlator eq \eqref{eq:Ambitwistor-JJO} is as follows 
    \begin{equation}
        \langle0|J^+(Z_1, W_1)J^+(Z_2, W_2)O_2(Z_3, W_3)|0\rangle = \int \prod_{i,j=1}^3 dc_{ij} ~e^{i\sum_{i,j=1}^{3} c_{ij} Z_i\cdot W_j} \frac{c_{13}c_{23}c_{31}c_{32}}{|c_{12}c_{23}c_{31}-c_{13}c_{21}c_{32}|^3}.
    \end{equation}
    The Penrose transform (after dropping the diagonal terms) would then simply be 
    \begin{align}
        \notag \braket{JJO} = \int dc_{ij} \frac{d^2\lambda_i d^2\tilde\lambda_i}{\mathrm{Vol}(GL(1, \mathbb R))^6} &(c_{21}\lambda_{2\alpha} + c_{31}\lambda_{3\alpha})(c_{12}\tilde\lambda_{2\dot\alpha} + c_{13}\tilde\lambda_{3\dot\alpha}) (c_{12}\lambda_{1\beta} + c_{32}\lambda_{3\beta})\\\notag&(c_{21}\tilde\lambda_{1\dot\beta} +  c_{23}\tilde\lambda_{3\dot\beta})~e^{i\sum_{i,j=1}^{3} c_{ij} \lambda_i \cdot x_{ij}\cdot \tilde\lambda_j} \frac{c_{13}c_{23}c_{31}c_{32}}{|c_{12}c_{23}c_{31}-c_{13}c_{21}c_{32}|^3}.
    \end{align}
    As was the case for two-points, the above integral, too, has multiple scaling redundancies, and one will have to re-parameterize to bring out these scaling directions.\footnote{For details on this variable change, and also the derivatives of the generating function for three points, we refer the reader to appendix \ref{app:penrose-details}.} Doing this, the correlator takes on the form  
     \begin{align}
         \notag\braket{J_{\alpha\dot\alpha}(x_1)J_{\beta\dot\beta}(x_2)O_2(x_3)} = \mathcal{N'}_{JJO}\int d^2\lambda_i d^2\tilde\lambda_i da_1 \frac{a_1}{|a_1-1|^3 }
        \left(\lambda _{2\alpha} + \lambda_{3\alpha}\right) \left( \tilde\lambda_{2\dot\alpha}+ \tilde\lambda_{3\dot\alpha}\right) 
        \left(\lambda _{1\beta} +  \lambda_{3\beta}\right) \\\left( \tilde\lambda_{1\dot\beta}+ a_1\tilde\lambda_{3\dot\beta}\right) \exp\left(i\sum_{(i,j)\ne (2,3)} \lambda_i\cdot x_{ij}\cdot \tilde\lambda_j + a_1 \,\lambda_2\cdot x_{23}\cdot \tilde\lambda_3  \right).
    \end{align}
    This can then be evaluated again as in the case of two points, by introducing sources, to give 
     \begin{equation}
     {
     \begin{aligned}
          \braket{J_{\alpha\dot\alpha}(x_1)J_{\beta\dot\beta}(x_2)O_2(x_3)} =\mathcal{N}_{JJO} \frac{\sigma^\mu_{\alpha\dot\alpha}\sigma^\mu_{\beta\dot\beta}}{x_{12}^2x_{23}^2x_{13}^2}\Bigg( &\left(\frac{1}{2}\frac{\eta_{\mu\nu}}{x_{12}^2} - \frac{x_{12\mu}x_{12\nu}}{x_{12}^4}  \right) \\
          & + \left(\frac{x_{13\mu}}{x_{13}^2} - \frac{x_{12\mu}}{x_{12}^2}\right)\left(\frac{x_{23\nu}}{x_{23}^2} - \frac{x_{21\nu}}{x_{21}^2}\right)\Bigg),   
     \end{aligned}
     }
    \end{equation}
    where \(\mathcal{N}_{JJO}\) is the overall normalisation, and it includes the contributions from all the integrals. We can see that it completely factorizes and separates from the tensor structure.\\

    \noindent\(\mathbf{\braket{JJJ}}\)\\
    The situation becomes considerably more nuanced when the CFT correlator admits multiple independent conformally invariant and conserved tensor structures. As an illustrative example, we consider the three-point function of conserved spin-1 currents, \(\braket{JJJ}\). This correlator admits two independent conserved and parity-even structures, which correspond holographically to the bulk \(F^2\) and \(F^3\) interaction vertices.
    
    For concreteness, we consider the Penrose transform of the ambitwistor correlator that has been matched with the \(F^2\) structure in momentum-space \cite{Bala:2026trw}. We choose to work with the all-minus helicity correlator in twistor space, which is given by
    \begin{align}
        \notag \braket{J^-&J^-J^-}_{F^2} =\\ &\int dc_{ij}\frac{\left(c_{13} \left(c_{22} c_{31}-c_{21} c_{32}\right)+c_{12} \left(c_{21} c_{33}-c_{23} c_{31}\right)+c_{11} \left(c_{23} c_{32}-c_{22} c_{33}\right)\right){}^2}{\left(c_{12} c_{23} c_{31}-c_{13} c_{21} c_{32}\right)|c_{12} c_{23} c_{31}-c_{13} c_{21} c_{32}|}\exp\left(ic_{ij}Z_i\cdot W_j \right).
    \end{align}
    After following the same steps as in the previous examples, we end up with the following expression for the position-space correlator
    \begin{align}\label{LessDiv}
        \braket{JJJ} = \notag\mathcal{N} \int  d^2\lambda_i d^2\tilde\lambda_i\, \int da_1\, \frac{(1+a_1)^2}{(1-a_1)|1-a_1|}\,\lambda_{1\alpha}\lambda_{1\dot\alpha}\lambda_{2\beta}\lambda_{2\dot\beta}\lambda_{3\gamma}\lambda_{3\dot\gamma}\\\exp\left(i\sum_{(i,j)\ne (2,3)} \lambda_i\cdot x_{ij}\cdot \tilde\lambda_j + a_1 \,\lambda_2\cdot x_{23}\cdot \tilde\lambda_3  \right).
    \end{align}
    We now switch the \(a_1\) and spinor integrals for convenience, and perform the \(\lambda\), \(\tilde\lambda\) integrals, which results in the following expression
    \begin{align}
       \notag \mathcal{N} \int da_1\,\frac{(1+a_1)^2}{|1-a_1|^7}\Bigg(a_1 \left(\frac{5\langle JJJ\rangle_{F^2}}{6} - \frac{\langle JJJ\rangle_{F^3}}{3}  \right) + \left(\frac{5\langle JJJ\rangle_{F^2}}{24} + \frac{\langle JJJ\rangle_{F^3}}{24} + \frac{\langle JJJ\rangle_{\epsilon}}{4}\right)\\ + \,a_1^2\left(\frac{5\langle JJJ\rangle_{F^2}}{24} + \frac{\langle JJJ\rangle_{F^3}}{24} - \frac{\langle JJJ\rangle_{\epsilon}}{4}\right)\Bigg),\label{eq:JJJ-naive}
    \end{align}
    which contains all the allowed position-space tensor structures. Here \(\langle JJJ\rangle_{F^2}\) and \(\langle JJJ\rangle_{F^3}\) are the two parity-even conserved structures corresponding to \(F^2\) and \(F^3\) interaction in the bulk  \cite{ Freedman:1998tz,Lee:2023qqx},
    \begin{align}
        \langle JJJ\rangle_{F^2} &= \frac{1}{x_{12}^4x_{23}^4x_{13}^4}\left(\frac{6}{5} V_1 V_2 V_3 + V_1 H_{23} + V_2 H_{13} + V_3 H_{12}\right),\\
        \langle JJJ\rangle_{F^3} &= \frac{1}{x_{12}^4x_{23}^4x_{13}^4}\left(6 V_1 V_2 V_3 + V_1 H_{23} + V_2 H_{13} + V_3 H_{12}\right),
    \end{align}
    with \(V_i\) and \(H_{ij}\) being the standard conformal building blocks \cite{Costa:2011mg} given as 
    \begin{align}
       \notag&V_{i,j,k~\alpha \dot\alpha} = \frac{x_{ij}^2 x_{ik~\alpha \dot{\alpha }}-x_{ ik}^2 x_{ ij ~\alpha \dot{\alpha }}}{x_{ jk}^2},~~H_{i,j~\alpha\dot\alpha \beta\dot\beta} = 2 \left(x_{ij}^2 \epsilon _{\alpha \beta } \epsilon _{\dot{\alpha }\dot{\beta }} + x_{ij~ \alpha \dot{\alpha }} x_{ji~ \beta \dot{\beta }}\right)\\
       &V_1 = V_{1,2,3},~ V_2 = V_{2,3,1},~ V_3 = V_{3,1,2},
    \end{align}
    and   
    \begin{equation}
        \langle JJJ\rangle_{\epsilon} = \frac{1}{x_{12}^4x_{23}^4x_{13}^4}\epsilon(\tilde x_{12}, \tilde z_1, \tilde z_2, z_3)  ,
    \end{equation}
    is the parity-odd conserved structure for \(\braket{JJJ}\) \cite{Costa:2011mg}. While we started with a structure that corresponds to the \(F^2\) structure in momentum-space, the Penrose transform has generated the full space of allowed conformally invariant and conserved tensor structures, including the parity-odd structure.

    A similar computation for \(\braket{TJJ}\) gives the following expression
    \begin{align}\label{eq:tjj-penrose}
       \mathcal{N} \int da_1\,\frac{(1+a_1)^2}{|1-a_1|^9}\Bigg(6\, a_1^2  \langle TJJ\rangle_{fb} + a_1 \left(\frac{\langle TJJ\rangle_{ff}}{4} + \langle TJJ\rangle_{\epsilon}\right) + \,a_1^3\left(\frac{\langle TJJ\rangle_{ff}}{4} - \langle TJJ\rangle_{\epsilon}\right)\Bigg)
    \end{align}
    which once again contains not only the free fermion and free boson basis structures \cite{Hartman:2016dxc,Erdmenger:1996yc}, but also the parity-odd structure.
    
   The \(a_1\) integral in eq \eqref{eq:JJJ-naive} is a divergent integral. In principle, the \(a_1\) integral in eq \eqref{LessDiv} should be performed first and then the spinor integrals, but this way of doing the integrals gets complicated, since the exponential structure will not be present anymore. In eq \eqref{LessDiv} the \(a_1\) integral already has a divergence at \(a_1 = 1\) even before performing the spinor integrals, and therefore the conditions required for exchanging the order of integration (e.g., absolute convergence or the hypotheses of Fubini-Tonelli) are not satisfied. Attempting to evaluate the \(a_1\) integral in eq \eqref{eq:JJJ-naive} using the conventional regularization techniques  (such as $\epsilon$ cutoff, Principal Value,  etc.) results in zero. To get the expected position-space expression, we therefore need to regulate the divergent integral appropriately. Below, we discuss a technique to regularize the integral.

    \subsubsection*{Regularizing the Integrals}
    Since the conventional regularization methods evaluate the integral to zero, we instead adopt the more general Epstein--Glaser framework\footnote{The Epstein--Glaser framework is formulated primarily as a systematic procedure for extending distributions that are singular at a point, with its original application being the construction of time-ordered products in causal perturbation theory. Here we use the same distributional extension idea to regularize the singular integral at hand.} \cite{Bresser:1999af, Gracia-Bondia:2000rpe}, which systematically characterizes all possible regularizations. Briefly put, the idea is to remove the divergent part of the Taylor expansion at the singular point. For concreteness, let us consider an integral of the following form, which is similar to the integral we have in hand
    \begin{equation}
        \int_{-\infty}^{\infty} dx\,\frac{\phi(x)}{|x|^k}.
    \end{equation}
    The singular behavior near \(x=0\) is completely determined by the first \(k\) terms in the Taylor expansion of \(\phi(x)\) around 0
    \begin{equation}
        \phi(x)=\sum_{n=0}^{k-1}\frac{\phi^{(n)}(0)}{n!}x^n+\mathcal{O}(x^k).
    \end{equation}
    Therefore, to get a finite value out of the integral, we need to remove the divergent Taylor terms, which leads to
    \begin{equation}\label{eq:regularization-Iw}
         I_w  = \int dx\, \frac{1}{|x|^k}\left(\phi(x) - w(x)\sum_{n=0}^{k-1}\frac{\phi^{(n)}(0)}{n!}x^n\right),
    \end{equation}
    where \(w(x)\) is a weight function that satisfies the following properties
    \begin{equation}\label{w(x)-properties}
        w(0) = 1, ~~~w^{(n)}(0) = 0~\forall n<k,~~~\lim_{x\to \infty} w(x) = 0 .
    \end{equation}
    The weight function \(w(x)\) is required because the highest-order subtraction term is proportional to \(x^{k-1}\). After multiplication by \(1/|x|^k\), this term behaves as \(1/|x|\), whose integral is logarithmically divergent at large \(|x|\). Thus, \(w(x)\) prevents the subtraction from introducing artificial divergence at infinity.
    
    Notice that the choice of weight function is arbitrary; hence there is an inherent ambiguity in the above regularization method. Different choices of \(w(x)\) adhering to the conditions in eq \eqref{w(x)-properties} give different finite values. However, the difference between two finite values will be\footnote{This is a well-defined integral since we have \(w^{(n)}(0) = 0~\forall n<k\) and therefore the lowest order term in the Taylor expansion at \(x=0\) is \(x^k\)}
    \begin{align}\label{eq:weight-diff}
        I_{w_2}-I_{w_1}
        &= \sum_{n<k}\phi^{(n)}(0)\int dx\,\frac{w_1(x)-w_2(x)}{n!\,|x|^k}x^n = \sum_{n<k}\phi^{(n)}(0) c_n.
    \end{align}
    Thus the ambiguity in regularization completely depends on a finite set of constants \(c_n\).  Restricting to symmetric regularization\footnote{Equation eq \eqref{eq:regularization-Iw} can be seen as regularizing the singular kernel \(\frac{1}{|x|^k}\) to give a finite value for a generic test function \(\phi\). Since the kernel is even under \(x\to -x\), we restrict to prescriptions that preserve this symmetry, which is the same as choosing an even \(w(x)\).} all odd \(c_n\) to vanish, so only the even coefficients remain.
    
   We now return to the $\langle JJJ\rangle$ integral in eq \eqref{eq:JJJ-naive}. The divergence is of degree seven, so in principle we should subtract the Taylor expansion of the numerator through order six in $(a_1-1)$. However, the numerator is only a polynomial of degree four. Its Taylor expansion around $a_1=1$ therefore contains no terms of order five or six. Consequently, there is no highest-order subtraction term that could introduce a new logarithmic divergence at large $a_1$. We may therefore choose the simple weight function $w(a_1)=1$ in eq \eqref{eq:regularization-Iw}. With this choice, the subtraction removes the entire numerator, since the numerator is exactly equal to its Taylor expansion. The regularized integrand for \(w=1\) thus vanishes identically, and the resulting finite part is zero:
    \begin{equation}
        I_{w=1}=0.
    \end{equation}
   The condition eq \eqref{eq:weight-diff} then states that for a general weight function, the finite value must be of the form    
   \begin{equation}\label{eq:regularized-jjj}
        I_{\mathrm{w}} = \langle\,JJJ\,\rangle_{F^2}\left(5c_0+\tfrac{85}{6}c_2+5c_4\right) + \langle\,JJJ\,\rangle_{F^3}\left(-c_0-\tfrac{13}{6}c_2+c_4\right)  + \langle\,JJJ\,\rangle_{\epsilon}\left(-6c_2-6c_4\right). 
    \end{equation}
    
    Knowing that the twistor ansatz goes to \(F^2\) structure in momentum-space \cite{Bala:2026trw}, we choose
    \begin{equation}
        c_0 =  \frac{19 \alpha }{40},~~c_2 =  -\frac{3 \alpha}{20} ,~~c_4 =  \frac{3 \alpha }{20},
    \end{equation}
    to obtain the expected position-space correlator
    \begin{equation}
        I_{\mathrm{reg}} = \alpha \braket{JJJ}_{F^2}.
    \end{equation}
     One can do a similar regularization for the divergent integrals in the computation of \(\braket{TJJ}\) in eq \eqref{eq:tjj-penrose} too, to fix the position-space correlator with the knowledge of the momentum-space structure.

    
    Thus, we see that the Penrose transform eq \eqref{eq:Ambi-penrose-spinning} of the ambitwistor correlators leads to the correct position-space result, subject to appropriate regularization. While the Grassmannian connection with ambitwistors guided us in the correct regularization, it is important to establish an independent understanding of the same.

    So far, we have presented the ambitwistor correlators and their Penrose transform. Given the connection to the Helicity-Basis Grassmannian, we now ask if a similar connection can be made to the covariant Grassmannian. Such a connection is indeed possible, but it will not lead to the ambitwistor space, but rather to the massive twistor space, which has manifest \(SL(2,\mathbb{R})\) covariance. Its construction and Penrose transform for these correlators will be the focus of our next section.

\section{Massive Twistors for CFT\(_4\) Correlators }\label{sec:SL2RTwistors}
In the off-shell setup, the massive twistor can be obtained starting with the off-shell spinor-helicity
$p_{\alpha\dot\alpha}=\lambda_{I\alpha}\tilde\lambda^{I}_{\dot\alpha}$, and then performing a half-Fourier transform on either $\lambda_\alpha^I$ or $\widetilde\lambda_{\dot{\alpha}}^I$, which results in
\begin{align}\label{SHTwistor}
(\lambda_\alpha^I,\widetilde\lambda_{\dot{\alpha}}^I)\;\xrightarrow{HFT}&\;Z^{I(A)}=(\lambda_\alpha^I,\widetilde\mu_{\dot{\alpha}}^I),\notag\\
(\lambda_\alpha^I,\widetilde\lambda_{\dot{\alpha}}^I)\;\xrightarrow{HFT}&\;W^I_{(A)}=(\mu_\alpha^I,\widetilde\lambda_{\dot{\alpha}}^I).
\end{align}
We define the massive twistor as $Z^{I(A)}$, which is also known as the massive twistor in the literature \cite{Albonico:2022pmd}. Massive twistors are defined to be the fundamental representation of the conformal group and the little group for the massive particle, which will be $SL(4,\mathbb R) \times SL(2,\mathbb R)$ in our context. The index $A$ represent $SL(4,\mathbb R)$, while index $I$ represent $SL(2,\mathbb R)$, which ensures a manifest little-group covariance. Note that $W$ is the Fourier conjugate variable of $Z$. Since $SL(4,\mathbb R)$ has only a fundamental representation, working with either $Z$ or $W$ is sufficient.

\textbf{Note:} While finalizing the manuscript, we became aware of \cite{S:2026qwn}. The following section contains some overlapping results with their work.

\subsection{Conformal Generators}
The conformal generators for massive twistors can be compactly written as
\begin{align}\label{SL2Generator}
T_A^B=Z^{I(B)}\frac{\partial}{\partial Z^{I(A)}}-\frac{1}{4}\delta_A^B Z\cdot\frac{\partial}{\partial Z},
\end{align}
which satisfy the $SL(4,\mathbb R)$ algebra
\begin{align}\label{SL2Algebra}
[T_A^B,T_C^D]=\delta_C^BT_A^D-\delta_A^DT_C^B.
\end{align}
However, the spinor helicity representation of the off-shell momenta eq \eqref{SLSH} has a little-group redundancy $SL(2,\mathbb{R}) \times GL(1, \mathbb R)$ and consequently, the conformal generators in twistor space also inherit this. The action of the little group generators is given by
\begin{align}\label{SL2Little}
SL(2,\mathbb{R}):&\qquad S^I_J=\frac{1}{2}Z^{I(A)}\frac{\partial}{\partial Z^{J(A)}} - \frac{1}{4}\delta_J^I Z^{K(A)}\frac{\partial}{\partial Z^{K(A)}}+ \hat{S}^I_J,\notag\\
GL(1,\mathbb{R}):&\qquad G=\frac{1}{2}Z^{I(A)}\frac{\partial}{\partial Z^{I(A)}}.
\end{align}
$\hat S^I_J$ represents the infinitesimal matrix representation, where its value depends on different representations of $SL(2,\mathbb{R})$, on which it acts. One can in principle construct the correlation functions in twistor space by solving eq \eqref{SL2Generator} and eq \eqref{SL2Little}. However, we find a more efficient way to obtain the correlators for $SL(2,\R)$ representation using the covariant Grassmannian formulation. 

\subsection{From Covariant Grassmannian to $SL(2,\mathbb{R})$ Twistors}\label{SL2Twistors}

We now illustrate the twistor space construction for correlators, starting from their Grassmannian analogues. The Grassmannian representation for $CFT_4$ correlators is given by \cite{Bala:2026trw}
\begin{align}\label{GrassmannianNPoint}
&\Psi_n^{\{I_1,\cdots\},\cdots,\{I_n,\cdots\} }(\Lambda,\widetilde \Lambda) \notag\\
&=\int\frac{\dd C~\dd\tilde C}{\Vol(GL(n))^2}
\delta(\tilde C\cdot\Omega\cdot C^T)\,
\delta(\tilde C^T\cdot\Omega\cdot\Lambda)\,
\delta(C^T\cdot\Omega\cdot\tilde{\Lambda})\,
\mathcal{A}_n^{\{I_1,\cdots\},\cdots,\{I_n,\cdots\} }(C,\tilde C),
\end{align}
where $\mathcal{A}_n(C,\widetilde C)$ is a function that depends on the $n\cross n$ minors of $\tilde C$ and $C$.

We now perform a half-Fourier transform with respect to $\widetilde\Lambda$, which takes the correlator to its twistor-space version
\begin{align}
&=\int\frac{dC\;d\widetilde C}{\mathrm{Vol}(GL(n))}\int d\widetilde\Lambda\;e^{i\widetilde M^T\cdot\Omega\cdot\widetilde\Lambda}\int d\widetilde\chi \;e^{i\widetilde\chi\cdot C^T\cdot\Omega\cdot\widetilde\Lambda}\;\delta(\widetilde C^T\cdot\Omega\cdot\Lambda)\;\delta(\widetilde C^T\cdot\Omega\cdot C)\; \mathcal{A}_n^{\{I_1,\cdots\},\cdots,\{I_n,\cdots\} }(C,\tilde C).
\end{align}
At this stage, we integrate out $\widetilde\Lambda$, which results in
\begin{align}
&=\int\frac{dC\;d\widetilde C}{\mathrm{Vol}(GL(n))^2}\;\int d\widetilde\chi\;\delta(\widetilde M^T+\widetilde\chi\cdot C^T)\delta(\widetilde C^T\cdot\Omega\cdot\Lambda)\;\delta(\widetilde C^T\cdot\Omega\cdot C)\; \mathcal{A}_n^{\{I_1,\cdots\},\cdots,\{I_n,\cdots\} }(C,\tilde C).
\end{align}
The $\widetilde\chi$ integral localizes $\widetilde M^T=-\widetilde\chi\cdot C^T$, which, under the symplectic orthogonality, can be expressed as $\widetilde M^T\cdot\Omega\cdot\widetilde C=0$. Thus, the above expression can be rewritten as
\begin{align}
&\hat{\psi}_n^{\{I_1,\cdots\},\cdots,\{I_n,\cdots\} }({Z})\notag \\ & =\int\frac{dC\;d\widetilde C}{\mathrm{Vol}(GL(n))^2}\;\delta(\widetilde C^T\cdot\Omega\cdot \Lambda)\;\delta(\widetilde C^T\cdot\Omega\cdot \widetilde M)\;\delta(\widetilde C^T\cdot\Omega\cdot C)\; \mathcal{A}_n^{\{I_1,\cdots\},\cdots,\{I_n,\cdots\} }(C,\tilde C),
\end{align}
which can be recast into
\begin{align}\label{SL2TwistorNPoint}
\boxed{\hat{\psi}_n^{\{I_1,\cdots\},\cdots,\{I_n,\cdots\} }({Z})=\int\frac{dC\;d\widetilde C}{\mathrm{Vol}(GL(n))^2}\;\delta^{4, n}(\widetilde C^T\cdot\Omega\cdot Z)\delta(\widetilde C^T\cdot\Omega\cdot C) \mathcal{A}_n^{\{I_1,\cdots\},\cdots,\{I_n,\cdots\} }(C,\tilde C).}
\end{align}
Thus, we observe that the Grassmannian formulation in eq \eqref{GrassmannianNPoint} renders a simple expression for $n$-point correlators in twistor space in eq \eqref{SL2TwistorNPoint}. Let us now present the twistor-space correlators to illustrate this point.

\subsection{Correlators in $SL(2,\mathbb{R})$ Twistors}
We present a few two- and three-point correlation functions for $SL(2,\R)$ twistors, starting with the former. The two-point function of arbitrary spin-$s$ operators takes the simple form
\begin{align}\label{SL2Twistor2Point}
\boxed{\hat{\psi}_2^{\{I_1,\cdots\},\{I_2,\cdots\} }({Z})=\int\frac{dC\;d\widetilde C}{\mathrm{Vol}(GL(2))^2}\;\delta^{4, 2}(\widetilde C^T\cdot\Omega\cdot Z)\delta(\widetilde C^T\cdot\Omega\cdot C) \mathcal{A}_2^{\{I_1,\cdots\},\{I_2,\cdots\} }(C,\tilde C),}
\end{align}
where the expressions for $\mathcal{A}_2(C,\widetilde C)$ are presented in \cite{Bala:2026trw}. For example, the two-point function of conserved spin-1 currents is
\begin{align}
\mathcal{A}_2^{(I_1,J_1),(I_2,J_2)}(C,\widetilde C)=\frac{(1^{(I_1}2^{(I_2})(\tilde{1}^{I_2)}\tilde{2}^{J_2)})}{(1_M2_N)(\tilde{1}^M\tilde{2}^N)}.
\end{align}
Subject to the symplectic orthogonality $\widetilde C^T\cdot\Omega\cdot C$ and $GL(2)$ gauge-fixing of $\widetilde C$ in eq \eqref{SL2Twistor2Point}, the remaining $\widetilde C$ elements are nothing but the Schwinger parameters in the twistor space.

The three-point functions follow a similar suit. Restricting eq \eqref{SL2TwistorNPoint} to three-point functions, the twistor-space result takes the following form
\begin{align}
\boxed{\hat{\psi}_3^{\{I_1,\cdots\},\cdots,\{I_3,\cdots\} }({Z})=\int\frac{dC\;d\widetilde C}{\mathrm{Vol}(GL(3))^2}\;\delta^{4, 3}(\widetilde C^T\cdot\Omega\cdot Z)\delta(\widetilde C^T\cdot\Omega\cdot C) \mathcal{A}_3^{\{I_1,\cdots\},\cdots,\{I_3,\cdots\} }(C,\tilde C),}
\end{align}
where the expressions for $\mathcal{A}_3(C,\widetilde C)$ are presented in \cite{Bala:2026trw}. For example, in $\langle O_2O_2O_2\rangle$\footnote{A different expression of the scalar three-point correlator is presented in \cite{CarrilloGonzalez:2026eum} for massive twistors. While their expression does not enjoy the manifest $SL(2,\R)$ little-group covariance, it is apparent in our case. Moreover, our expression goes to the correct momentum-space result after performing an inverse half-Fourier transform. It would be interesting to connect these two pictures.} 
\begin{align}
\mathcal{A}_{3}=\frac{\Sgn(\mathcal{K})}{\mathcal{K}},
\end{align}
where $\Block= (1^I 1^J 2^K) (\ctil{1}_I \ctil{1}_J \ctil{2}_K)$.


Having developed the $SL(2,\R)$ twistors starting from the Grassmannian formulation of momentum-space, we now want to obtain their position-space analogue. To that end, we will now derive the Penrose transform.
\subsection{Penrose transform for $SL(2,\mathbb{R})$ Twistors}\label{SL2RPenrose}
In this section, we will formally develop the Penrose transform for $SL(2,\mathbb{R})$ twistors before obtaining some position-space correlators to illustrate this machinery.

\subsubsection{Derivation of Penrose Transform}
We start with the basic definition of the Fourier transform for conserved spin-1 currents
\begin{align}
J_{\alpha\dot{\alpha}}(x)&=\int d^4p\;e^{ip\cdot x} J_{\alpha\dot{\alpha}}(p)=\int \frac{d^4\lambda d^4\widetilde\lambda}{\textrm{Vol}(SL(2,\mathbb R))\textrm{Vol}(GL(1,\mathbb R))}\;e^{i\lambda_\beta\cdot\widetilde\lambda_{\dot{\beta}}x^{\beta\dot{\beta}}}\epsilon_{\gamma\alpha}\epsilon_{\dot{\gamma}\dot{\alpha}}\;J^{\gamma\dot{\gamma}}(p).
\end{align}
The above equation can be rewritten in the following form
\begin{align}
J_{\alpha\dot{\alpha}}(x)&=\int \frac{d^4\lambda d^4\widetilde\lambda}{\textrm{Vol}(GL(2,\mathbb R))}\;e^{i\lambda_\beta\cdot\widetilde\lambda_{\dot{\beta}}x^{\beta\dot{\beta}}}\lambda_{\alpha,I}\widetilde\lambda_{\dot{\alpha},J}\hat{J}^{(IJ)}(p),
\end{align}
where we have used the identity $p^2\epsilon_{\alpha\gamma}\epsilon_{\dot{\alpha}\dot{\gamma}}=p_{\alpha\dot{\alpha}}p_{\gamma\dot{\gamma}}-p_{\alpha\dot{\gamma}}p_{\gamma\dot{\alpha}}$. The contribution from first term in the above expression drops off due to the conservation equation $p_{\gamma\dot{\gamma}}J^{\gamma\dot{\gamma}}(p)=0$. Moreover, we define $\hat{J}^{(IJ)}(p)\equiv \zeta_\gamma^I\widetilde\zeta_{\dot{\gamma}}^J \frac{J^{\gamma\dot{\gamma}}(p)}{p}$.\footnote{Here, the polarization spinors are defined as $\zeta_{\alpha}^I=\frac{\lambda_\alpha^I}{\sqrt{p}}\;\textrm{and}\;\widetilde\zeta_{\dot{\alpha}}^I=\frac{\widetilde\lambda_{\dot{\alpha}}^I}{\sqrt{p}}$. Moreover, the current has been rescaled by $p^{-1}$ to ensure that $\hat{J}$ scales as $\Delta=2$ operator.} Expressing the current in terms of its twistor-space representative, we obtain
\begin{align}
J_{\alpha\dot{\alpha}}(x)&=\int \frac{d^4\lambda d^4\widetilde\lambda}{\textrm{Vol}(GL(2,\mathbb R))}\;e^{i\lambda_\beta\cdot\widetilde\lambda_{\dot{\beta}}x^{\beta\dot{\beta}}}\lambda_{\alpha,I}\widetilde\lambda_{\dot{\alpha},J}\int d^4\widetilde\mu\;e^{i\widetilde\mu\cdot\widetilde\lambda}\hat{J}^{(IJ)}(\lambda,\widetilde\mu).
\end{align}
Finally, performing the $\widetilde\lambda$ integral and using integration by parts, we obtain the final result for the Penrose transform for conserved spin-1 currents
\begin{align}
J_{\alpha\dot{\alpha}}(x)&=\int \frac{d^4\lambda}{\textrm{Vol}(GL(2,\mathbb R))}\;\lambda_{\alpha,I}\frac{\partial}{\partial\tilde\mu^{\dot{\alpha},J}}\hat{J}^{(IJ)}(\lambda,\tilde\mu)\Big|_X,
\end{align}
subject to the incidence relation
\begin{align}
X:\quad \widetilde\mu^{\dot{\kappa},K}=x^{\kappa\dot{\kappa}}\lambda^K_\kappa. 
\end{align}
One can similarly establish the Penrose transform for conserved integer spin-$s$ currents
\begin{align}\label{SL2RPenroseTransform}
\boxed{J_s^{\alpha_1\cdots\alpha_s,\dot{\alpha}_1\cdots\dot{\alpha}_s}(x)=\int\frac{d^4\lambda d^4\widetilde\mu}{\textrm{Vol}(GL(2,\mathbb R))}\prod_{i=1}^s\prod_{j=1}^s\lambda_{I_i}^{\alpha_i}\frac{\partial}{\partial\widetilde\mu_{\dot{\alpha}_j}^{J_j}}\hat{J}^{(I_1\cdots I_s J_1\cdots J_s)}(\lambda,\widetilde\mu)\Big\vert_X.}
\end{align}
Let us now obtain position-space correlators, starting from their twistor-space representatives, using the Penrose transform eq \eqref{SL2RPenroseTransform}.

\subsubsection{Position-Space Results}
We now derive the position-space correlators using the Penrose transform, starting with two-point functions.
\subsubsection*{Two-Point Functions}
Consider the scalar two-point function $\langle O_2O_2\rangle$. The Grassmannian representative for this correlator is simply $f(C,\widetilde C)=1$. Using the relation between Grassmannian and twistor space, eq \eqref{SL2TwistorNPoint}, we obtain the twistor-space expression
\begin{align}
\langle0|O_2(Z_1)O_2(Z_2)|0\rangle&=\int \frac{d\widetilde C}{\textrm{Vol}(GL(2))}\;\delta^{2,4}[\widetilde C_{ij,I}Z^{I,(A)}_j].
\end{align}
Restricting the Penrose transform to the scalar case, we obtain
\begin{align}
\langle0|O_2(x_1)O_2(x_2)|0\rangle&=\int\prod_{k=1}^2\frac{d\lambda_k^K}{\textrm{Vol}(SL(2,\mathbb{R}))}\int \frac{d\widetilde C}{\textrm{Vol}(GL(2))}\;\delta^{2,4}[\widetilde C_{ij,I}Z^{I,(A)}_j]\Big|_{\widetilde\mu^{\dot{\kappa},K}_k=x^{\kappa\dot{\kappa}}\lambda^K_{k,\kappa}}.
\end{align}
We now gauge fix $\widetilde C$ using the $GL(2)$ redundancy
\begin{align}\label{GL2}
\widetilde C=\begin{pmatrix}
1 & 0 & a & b \\
0 & 1 & c & d
\end{pmatrix}=(1_{2\cross2}|\widetilde c_{2\cross2}),
\end{align}
which allows us to rewrite the above expression as
\begin{align}
\langle0|O_2(x_1)O_2(x_2)|0\rangle
&=\int\prod_{k=1}^2\frac{d\lambda_k^K}{\textrm{Vol}(SL(2,\mathbb{R}))}\int d\widetilde c\;\delta^{2,2}(\lambda_1^{a,I}+c_J^I\lambda_2^{a,J})\delta^{2,2}(\widetilde\mu_1^{\dot{a},I}+c_J^I\widetilde\mu_2^{\dot{a},J})\Big|_{\widetilde\mu^{\dot{\kappa},K}_k=x^{\kappa\dot{\kappa}}\lambda^K_{k,\kappa}}.
\end{align}
Subject to the incidence relation and integrating $\lambda_1$, the above equation becomes
\begin{align}
\langle0|O_2(x_1)O_2(x_2)|0\rangle
&=\int\frac{d\lambda_2}{\textrm{Vol}(SL(2,\mathbb{R}))}\int d\widetilde c\;\delta^{2,2}(\widetilde c_J^I(x_{12,a}^{\dot{a}})\lambda_2^{a,I}).
\end{align}
Finally, the above integral can be written as
\begin{align}
\langle0|O_2(x_1)O_2(x_2)|0\rangle&=\frac{1}{\textrm{Det}(x_{12})^2}\int\frac{d\lambda_2^K}{\textrm{Vol}(SL(2,\mathbb{R}))}\frac{1}{\textrm{Det}(\lambda_2)}\int d\widetilde c\;\delta^{2,2}(\widetilde c_J^I).
\end{align}
The $\lambda$ integral is unity\footnote{By definition, the Haar measure is $\int\frac{d\lambda}{\textrm{Det}(\lambda)}=\textrm{Vol}(SL(2,\mathbb{R}))$.}, while the $\widetilde c$ integral is trivial. Thus, the final expression of the correlator boils down to the correct position-space two-point function
\begin{align}
\langle0|O_2(x_1)O_2(x_2)|0\rangle&=\frac{1}{x_{12}^4}.
\end{align}
We now move on to the spinning correlators. Consider spin-1 conserved currents, where their Grassmannian representative $f(C,\widetilde C)$ \cite{Bala:2026trw} results in the following twistor-space expression\footnote{Here, we have already used the symplectic orthogonality, which leads to the constraint $C=\widetilde C_\perp$. We will henceforth impose this condition in all subsequent computations.}
\begin{align}
\langle0|J^{I_1J_1}(Z_1)J^{I_2J_2}(Z_2)|0 \rangle&=\int\prod_{k=1}^2\frac{d\lambda_k^K}{\textrm{Vol}(SL(2,\mathbb{R}))}\int \frac{d\widetilde C}{\textrm{Vol}(GL(2))}\;\delta^{2,4}[\widetilde C_{ij,I}Z^{I,(A)}_j]\frac{C^{(I_1(I_2}C^{J_1)J_2)}}{\textrm{Det}(C)}\Big|_{C=\widetilde C_\perp},
\end{align}
and the position-space version can be obtained by using the Penrose transform eq \eqref{SL2RPenroseTransform}
\begin{align}
\langle0|J^{a\dot{a}}(x_1)J^{b\dot{b}}(x_2)|0 \rangle=&\int\prod_{k=1}^2\frac{d\lambda_k^K}{\textrm{Vol}(SL(2,\mathbb{R}))}\lambda_{1I_1}^a\lambda_{2I_2}^b\frac{\partial}{\partial\widetilde\mu_{1\dot{a}}^{J_1}}\frac{\partial}{\partial\widetilde\mu_{2\dot{b}}^{J_2}}\notag\\
&\int \frac{d\widetilde C}{\textrm{Vol}(GL(2))}\;\delta^{2,4}[\widetilde C_{ij,I}Z^{I,(A)}_j]\frac{C^{(I_1(I_2}C^{J_1)J_2)}}{\textrm{Det}(C)}\Big|_{C=\widetilde C_\perp}\Big|_{\widetilde\mu^{\dot{\kappa},K}_k=x^{\kappa\dot{\kappa}}\lambda^K_{k,\kappa}}.
\end{align}
Using the gauge-fixing eq \eqref{GL2} as above, we express the delta functions as some Schwinger integrals. The action of derivatives, the subsequent imposition of the incidence relation, and integrating out $\lambda_1^{k,K}$ lead to the following expression
\begin{align}
\langle0|J^{a\dot{a}}(x_1)J^{b\dot{b}}(x_2)|0 \rangle=&\int\frac{d\lambda_2^K}{\textrm{Vol}(SL(2,\mathbb{R}))}\int d\widetilde c\;\widetilde c_{I_1}^P\widetilde c_{J_2}^K\frac{c^{(I_1(I_2}c^{J_1)J_2)}}{\textrm{Det}(c)}\Big|_{c=\widetilde c_\perp}\notag\\
&\int d\widetilde q\;q^{\dot{a}}_{J_1} \widetilde q^{\dot{b}}_K\lambda_{2P}^a\lambda_{2I_2}^b\exp( i\widetilde q_{\dot{m},M}\widetilde c_N^M x_{12,m}^{\dot{m}}\lambda_2^{m,N}).
\end{align}
The task now is to integrate by parts, by invoking appropriate functional derivatives that act on the exponential. In order to do so, we repackage the above equation as
\begin{align}
\langle0|J^{a\dot{a}}(x_1)J^{b\dot{b}}(x_2)|0 \rangle=&\int\frac{d\lambda_2^K}{\textrm{Vol}(SL(2,\mathbb{R}))}\int d\widetilde c\;\widetilde c_{I_1}^P\widetilde c_{J_2}^K\frac{c^{(I_1(I_2}c^{J_1)J_2)}}{\textrm{Det}(c)}\Big|_{c=\widetilde c_\perp}\notag\\
&\int d\widetilde q\;\widetilde q^{\dot{b}}_K\lambda_{2P}^a\lambda_{2I_2}^b\Bigg(\frac{x_{12}^{r\dot a}}{x_{12}^2} (\widetilde c_{J_1}^R)^{-1}\frac{\partial}{\partial \lambda_{2}^{rR}}\exp( i\widetilde q_{\dot{m},M}\widetilde c_N^M x_{12,m}^{\dot{m}}\lambda_2^{m,N})\Bigg).
\end{align}
Integrating by parts by repeating the same process and dropping the boundary terms, we obtain the following expression
\begin{align}
\langle0|J^{a\dot{a}}(x_1)J^{b\dot{b}}(x_2)|0 \rangle=&\int\frac{d\lambda_2^K}{\textrm{Vol}(SL(2,\mathbb{R}))}\int d\widetilde c\;\frac{c^{(I_1(I_2}c^{J_1)J_2)}}{\textrm{Det}(c)}\Big|_{c=\widetilde c_\perp}\delta(\widetilde c_N^M x_{12,m}^{\dot{m}}\lambda_2^{m,N})\notag\\
&\Big(\frac{x_{12}^{a\dot a}x_{12}^{b\dot b}}{x_{12}^4}\epsilon_{I_1J_1}\epsilon_{I_2J_2}+\frac{x_{12}^{a\dot b}x_{12}^{b\dot a}}{x_{12}^4} (\widetilde c_{I_1J_2})(\widetilde c_{I_2J_1})^{-1}\Big).
\end{align}
The first term in the above equation is zero by symmetry arguments, while the second term can be recast into the following form after doing the $\lambda_2$ integral
\begin{align}
\langle0|J^{a\dot{a}}(x_1)J^{b\dot{b}}(x_2)|0 \rangle=
&\frac{1}{\textrm{Det}(x_{12})^2}\Bigg(\frac{\epsilon^{ab}\epsilon^{\dot{a}\dot{b}}}{x_{12}^2}-\frac{x_{12}^{a\dot{a}}x_{12}^{b\dot{b}}}{x_{12}^4}\Bigg)\int d\widetilde c\;\frac{(\widetilde c_{I_1J_2})(\widetilde c_{I_2J_1})^{-1}c^{(I_1(I_2}c^{J_1)J_2)}}{\textrm{Det}(c)\textrm{Det}(\widetilde c)}\Big|_{c=\widetilde c_\perp}\delta(\widetilde c_N^M).
\end{align}
The $\widetilde c$ integral is a homogeneity-0 integral that results in a constant that can be absorbed in the normalization of the correlator. The remaining piece can be written in terms of the inversion tensor
\begin{align}
\langle0|J^{a\dot{a}}(x_1)J^{b\dot{b}}(x_2)|0 \rangle=\frac{\mathcal{N}_{JJ}\;(\sigma_\mu)^{a\dot{a}}(\sigma_\nu)^{b\dot{b}}}{x_{12}^6} \left(\eta_{\mu\nu} - 2\frac{x_{12\mu}x_{12\nu}}{x_{12}^2} \right),   
\end{align}
which results in the correct expression for the two-point correlator for spin-1 currents in position space. Similarly, one can show that the two-point function for arbitrary integer spin can be simply computed by iteratively doing integration by parts
\begin{align}
\langle0|J_s^{a_1\cdots a_s,\dot{a}_1\cdots\dot{a}_s}(x_1)J_s^{b_1\cdots b_s,\dot{b}_1\cdots\dot{b}_s}(x_2)|0 \rangle&=\frac{\mathcal{N}_{J_s J_s}}{x_{12}^4}\prod_{i=1}^s\Bigg(\frac{\epsilon^{a_ib_i}\epsilon^{\dot{a}_i\dot{b}_i}}{x_{12}^2}-\frac{x_{12}^{a_i\dot{a}_i}x_{12}^{b_i\dot{b}_i}}{x_{12}^4}\Bigg).
\end{align}
Let us now move to three-point functions.

\subsubsection*{Three-Point Functions}
We begin with the simple case of $\langle O_2O_2O_2\rangle$, whose Grassmannian expression is $f(C,\widetilde C)=\frac{\textrm{Sgn}(\mathcal{K})} {\mathcal{K}}$ \cite{Bala:2026trw}. Using eq \eqref{SL2TwistorNPoint}, we have the twistor-space correlator
\begin{align}
\langle0|O_2(Z_1)O_2(Z_2)O_2(Z_3)|0\rangle=\int \frac{d\widetilde C}{\textrm{Vol}(GL(3))}\;\delta^{3,4}[\widetilde C_{ij,I}Z^{I,(A)}_j]\frac{\textrm{Sgn}(\mathcal{K}(C,\widetilde C))}{\mathcal{K}(C,\widetilde C)}\Bigg|_{C=\widetilde C_\perp}.
\end{align}
The position-space expression of the same can be obtained by the Penrose transform eq \eqref{SL2RPenroseTransform}
\begin{align}
\langle0|O_2(x_1)O_2(x_2)O_2(x_3)|0\rangle=\int\prod_{k=1}^3\frac{d\lambda_k^K}{\textrm{Vol}(SL(2,\mathbb{R}))}\int \frac{d\widetilde C}{\textrm{Vol}(GL(3))}\;\delta^{3,4}[\widetilde C_{ij,I}Z^{I,(A)}_j]\frac{\textrm{Sgn}(\mathcal{K}(C,\widetilde C))}{\mathcal{K}(C,\widetilde C)}\Bigg|_{\substack{C=\widetilde C_\perp,\\
\widetilde\mu^{\dot{\kappa},K}_k=x^{\kappa\dot{\kappa}}\lambda^K_{k,\kappa}}}.
\end{align}
Gauge fixing $\widetilde C$ using the $GL(3)$ redundancy
\begin{align}\label{GL3}
\widetilde C=\begin{pmatrix}
1 & 0 & 0 & a & b & c\\
0 & 1 & 0 & d & e & f\\
0 & 0 & 1 & g & h& i
\end{pmatrix}=(1_{3\cross3}|\widetilde c_{3\cross3}),
\end{align}
and working component-wise (e.g. $\lambda_{i,I}^a$ etc.), along with the incidence relation results in
\begin{align}
\langle0|O_2(x_1)O_2(x_2)O_2(x_3)|0\rangle=&\int\prod_{k=1}^3\frac{d\lambda_k^K}{\textrm{Vol}(SL(2,\mathbb{R}))}\int {d\widetilde c}\;\frac{2(cgh-afh-bdi+aei)}{g}\notag\\
&\delta(\lambda_{1,2}^a-a\lambda_{2,1}^a+b\lambda_{3,2}^a-c\lambda_{3,1}^a)\delta(x_{1a}^{\dot{a}}\lambda_{1,2}^a-ax_{2a}^{\dot{a}}\lambda_{2,1}^a+bx_{3a}^{\dot{a}}\lambda_{3,2}^a-cx_{3a}^{\dot{a}}\lambda_{3,1}^a)\notag\\
&\delta(-\lambda_{1,1}^a-d\lambda_{2,1}^a+e\lambda_{3,2}^a-f\lambda_{3,1}^a)\delta(-x_{1a}^{\dot{a}}\lambda_{1,2}^a-dx_{2a}^{\dot{a}}\lambda_{2,1}^a+ex_{3a}^{\dot{a}}\lambda_{3,2}^a-fx_{3a}^{\dot{a}}\lambda_{3,1}^a)\notag\\
&\delta(\lambda_{2,2}^a-g\lambda_{2,1}^a+h\lambda_{3,2}^a-i\lambda_{3,1}^a)\delta(x_{2a}^{\dot{a}}\lambda_{2,2}^a-gx_{2a}^{\dot{a}}\lambda_{2,1}^a+hx_{3a}^{\dot{a}}\lambda_{3,2}^a-ix_{3a}^{\dot{a}}\lambda_{3,1}^a).
\end{align}
Integrating all the $\lambda_{i,I}$ then results in the following expression
\begin{align}
\langle0|O_2(x_1)O_2(x_2)O_2(x_3)|0\rangle=&\frac{1}{x_{12}^2x_{23}^2x_{31}^2}\int \frac{d\widetilde c}{\textrm{Vol}(SL(2,\mathbb{R}))^3}(cgh-afh-bdi+aei)^2.
\end{align}
One then takes note of the fact that the $\widetilde c$ integral is of homogeneity zero, and thus results in a constant that we can absorb in the normalization $\mathcal{N}_{OOO}$. The remaining result is the correct three-point function for scalars.

We now turn our attention to correlators involving spinning operators. Let us consider the $\langle JO_2O_2\rangle$ for simplicity, whose twistor-space correlator is given by
\begin{align}
\langle0|J^{(IJ)}(Z_1)O_2(Z_2)O_2(Z_3)|0\rangle=\int \frac{d\widetilde C}{\textrm{Vol}(GL(3))}\;\delta^{3,4}[\widetilde C_{ij,I}Z^{I,(A)}_j]\;(1^{(I} 2^K 3^L)(\widetilde 1^{J)} \widetilde 2_K \widetilde 3_L)\frac{\textrm{Sgn}(\mathcal{K}(C,\widetilde C))}{\mathcal{K}(C,\widetilde C)^2}\Bigg|_{C=\widetilde C_\perp},
\end{align}
and the position-space expression of the same can be obtained by the Penrose transform eq \eqref{SL2RPenroseTransform}
\begin{align}
\int\prod_{k=1}^3\frac{d\lambda_k^K}{\textrm{Vol}(SL(2,\mathbb{R}))}\;&\lambda_{1,I}^{a}\frac{\partial}{\partial\tilde\mu_{1\dot a}^{J}}\int \frac{d\widetilde C}{\textrm{Vol}(GL(3))}\;\delta^{3,4}[\widetilde C_{ij,I}Z^{I,(A)}_j](1^{(I} 2^K 3^L)(\widetilde 1^{J)} \widetilde 2_K \widetilde 3_L)\frac{\textrm{Sgn}(\mathcal{K}(C,\widetilde C))}{\mathcal{K}(C,\widetilde C)}\Bigg|_{\substack{C=\widetilde C_\perp,\\
\widetilde\mu^{\dot{\kappa},K}_k=x^{\kappa\dot{\kappa}}\lambda^K_{k,\kappa}}}.
\end{align}
We now gauge-fix using eq \eqref{GL3}, write the delta-functions as some Schwinger integrals, and act the derivatives on them. Imposing the incidence relation then results in
\begin{align}
\langle0|J^{a\dot{a}}(x_1)O_2(x_2)O_2(x_3)|0\rangle=&\int\prod_{k=1}^3\frac{d\lambda_k^K}{\textrm{Vol}(SL(2,\mathbb{R}))}\int d\widetilde c\;\widetilde q_i^{\dot a}\widetilde  c_{i1,J}\lambda_{1,I}^{a}\;\delta(\widetilde c_{ij,M}\lambda_j^{p,M})\notag\\
&\int d\widetilde q\;\exp(\widetilde q_{i,\dot m}\widetilde c_{ij,M}x_{j,m}^{\dot m}\lambda_j^{m,M})(1^{(I} 2^K 3^L)(\widetilde 1^{J)} \widetilde 2_K \widetilde 3_L)\frac{\textrm{Sgn}(\mathcal{K}(C,\widetilde C))}{\mathcal{K}(C,\widetilde C)}\Bigg|_{\substack{C=\widetilde C_\perp,\\
\widetilde\mu^{\dot{\kappa},K}_k=x^{\kappa\dot{\kappa}}\lambda^K_{k,\kappa}}}.
\end{align}
One now follows the same exercise as in the two-point spinning case: integrating by parts and using the delta-function integrals to obtain some position-space terms multiplied by some Schwinger integrals. We skip these simple (but lengthy) intermediate steps, which lead to the following result
\begin{align}
\langle0|J^{a\dot{a}}(x_1)O_2(x_2)O_2(x_3)|0\rangle=\frac{\mathcal{N}_{JOO}}{x_{12}^2x_{23}^2x_{31}^2}\Bigg(\frac{x_{12}^{a\dot a}}{x_{12}^2}+\frac{x_{31}^{a\dot a}}{x_{31}^2}\Bigg),
\end{align}
which is the correct position-space result for $\langle JO_2O_2\rangle$, up to the normalization constant $\mathcal{N}_{JOO}$ obtained from the $\widetilde c$ integral.

The other correlators can similarly be obtained by following this procedure. One just needs to be careful about the appropriate regularization in all spinning correlators (see subsection \ref{sec:Pen} for detailed commentary on this matter). Thus, we observe that our $SL(2,\R)$ twistor formulation eq \eqref{SL2TwistorNPoint} gives correct position-space correlators via the Penrose transform eq \eqref{SL2RPenroseTransform}.

\section{Conclusion and Discussion}\label{sec:Disc}
In this paper, we have developed the real twistor space for four-dimensional CFT correlators. We solve for the ambitwistor correlators using conformal Ward identities along with little-group covariance. Moreover, we present a complementary route to obtain these correlators, starting from the helicity-basis Grassmannian representatives. We then derive the Penrose transform for the ambitwistor formulation, which leads to the position-space correlators. We further show that the half-Fourier transform from the covariant Grassmannian leads to the massive twistor representation, which manifests the little-group covariance. The Penrose transform of these massive twistor expressions also results in their correct position-space counterparts.

There are several interesting avenues for further exploration, some of which are outlined below.
\subsection*{Supertwistors}
A natural extension of our construction is to incorporate supersymmetry through supertwistor variables. Since supertwistors furnish a linear realization of superconformal symmetry \cite{Bala:2025jbh,Mazumdar:2025egx,S:2026qwn}, it would be interesting to formulate superconformal correlators directly in twistor space. An especially important case is $\mathcal{N}=4$ SYM, where twistor methods have led to remarkable simplifications in scattering amplitudes. It would be interesting to explore whether a similar organization emerges for correlation functions.

\subsection*{Higher-point Functions}

While this work focused on two- and three-point correlators, it would be interesting to generalize this method to obtain higher-point correlators \cite{S:2026qwn, Abhayankar:2026}. An important question is whether the Schwinger-parameter representation developed here admits a natural extension to higher-point functions and whether it acquires factorization properties in twistor space. The connection to Grassmannian formulations suggests that such a structure may exist, particularly in light of the BCFW bridge due to the factorization properties already observed in four-point CFT$_3$ correlators \cite{Bala:2026lvw}.
Another promising direction is to construct higher-point functions via conformal bootstrap. A possible avenue is via developing the quadratic and quartic conformal Casimir operators in ambitwistor space and solving the corresponding equations to obtain conformal blocks \cite{Arundine:2026myr}.


\acknowledgments
We thank K.S. Dhruva for useful discussions. AB acknowledges a UGC-JRF fellowship. AAR acknowledges a CSIR-JRF fellowship. We acknowledge our debt to the people of India for their constant support of research in basic sciences.

\appendix
\vspace{20pt}
\noindent
{\Large\textbf{Appendix}}

\section{Review of 4d off-shell Spinor-Helicity}
In this appendix, we quickly review the off-shell spinor-helicity formalism and the symplectic Grassmannian in four dimensions presented in \cite{Bala:2026trw}.  The central observation in that work was the use of off-shell spinor-helicity variables, and setting the stage for Grassmannian formalism, which makes manifest the momentum conservation, conformal invariance, and little-group covariance simultaneously.

\subsection*{Off-shell Spinor-Helicity in $d=4$}\label{app:OffshellSH}
We work in Klein space, $\mathbb{R}^{2,2}$ throughout this paper, where the spinor 
helicity variables and twistors are real-valued since $SO(2,2)\cong SL(2,\mathbb{R})_L\times SL(2,\mathbb{R})_R$. Unlike the on-shell four-dimensional spinor-helicity formalism, the off-shell setup has an enlarged little group $SL(2,\R)\times GL(1,\R)$. The momentum vector can be traded for a pair of spinor variables. In the covariant spinor-helicity representation, the off-shell momentum can be written as
\begin{align}\label{SLSH}
p_{\alpha\dot\alpha}
= \lambda_{I\alpha}\tilde{\lambda}^{I}{}_{\dot\alpha}
= \lambda_{I\alpha}\tilde{\lambda}_{J\dot\alpha}\epsilon^{IJ},
\end{align}
where \(\alpha\) and \(\dot\alpha\) are the two real spinor indices of \(SL(2,\R)_L\) and \(SL(2,\R)_R\), while \(I\) is the little-group $SL(2,\R)$ index. The parametrization has eight real spinor components for the four real components of \(p_{\alpha\dot\alpha}\), and the four redundant degrees are precisely
\begin{equation}
\lambda_{I\alpha}\mapsto r\,\lambda_{I\alpha} \quad
\& \quad
\tilde{\lambda}^{I}{}_{\dot\alpha}\mapsto r^{-1}\tilde{\lambda}^{I}{}_{\dot\alpha},
\qquad
\lambda_{I\alpha}\mapsto S_I{}^{J}\lambda_{J\alpha},
\quad\&\quad
\tilde{\lambda}^{I}{}_{\dot\alpha}\mapsto (S^{-T})^I{}_{J}\tilde{\lambda}^{J}{}_{\dot\alpha}.
\end{equation}
The first pair is the $GL(1,\R)$ scaling, which removes one component. While the second pair is the $SL(2,\R)$ little-group rotation, which removes another three components. The corresponding helicity-basis representation has an \(SL(2,\R)\) index,\cite{Bala:2026trw} which is given as follows:
\begin{align}
    \xi^I_\alpha= \frac{\lambda_{\alpha}^{I}}{Det(\lambda)}, \qquad \bar\xi^I_{\dot\alpha}= \frac{\tilde\lambda_{\dot\alpha}^{I}}{Det(\tilde\lambda)} .
\end{align}

Alternatively, one can set up the off-shell spinor-helicity, in the same spirit as the massive spinor-helicity formalism, by introducing two pairs of spinors
\begin{align}\label{SH}
p_{\alpha\dot\alpha}=\lambda_{I\alpha}\tilde\lambda^{I}_{\dot\alpha}
=\lambda_\alpha \tilde\rho_{\dot\alpha}-\rho_\alpha \tilde\lambda_{\dot\alpha},
\end{align}
where $\lambda_\alpha^I=\{\lambda_\alpha,\rho_\alpha\}$ and $\tilde\lambda_{\dot\alpha}^I=\{\tilde\rho_{\dot\alpha}, \tilde\lambda_{\dot\alpha}\}$ are real spinors transforming 
in the fundamental representations of $SL(2,\mathbb{R})_L$ and $SL(2,\mathbb{R})_R$ 
respectively, with indices $\alpha,\dot\alpha = 1,2$. The corresponding helicity-basis representation resolves the little-group doublet into two component-level spinors. This form is useful when one wants to study individual helicity components, rather than fully covariant $SL(2,\R)$ multiplets. For integer spin, the three polarization vectors may be chosen as
\begin{equation}
\xi^{(+)}_{\alpha\dot\alpha}
=\frac{\lambda_\alpha\tilde{\lambda}_{\dot\alpha}}{p},
\qquad
\xi^{(-)}_{\alpha\dot\alpha}
=\frac{\rho_\alpha\tilde{\rho}_{\dot\alpha}}{p},
\qquad
\xi^{(0)}_{\alpha\dot\alpha}
=\frac{\rho_\alpha\tilde{\lambda}_{\dot\alpha}+\lambda_\alpha\tilde{\rho}_{\dot\alpha}}{p}.\label{eq:Helicity-basis-spinors}
\end{equation}
While for half-integer operators, one uses the following spinor polarizations
\begin{align}
\xi^{\alpha(+)}_L=\frac{\lambda^\alpha}{\sqrt{p}},\qquad \xi^{\alpha(-)}_L=\frac{\rho^\alpha}{\sqrt{p}},\qquad \xi^{\dot{\alpha}(+)}_R=\frac{\tilde{\lambda}^{\dot{\alpha}}}{\sqrt{p}},\qquad \xi^{\dot{\alpha}(-)}_R=\frac{\tilde{\rho}^{\dot{\alpha}}}{\sqrt{p}}.
\label{Helicity-basis-polarization-spinors}
\end{align}
This setup serves as the starting point for representing operators in the helicity basis by dotting the operators with the polarization basis.

\section{Solving for \(\mathcal{G}(c_{mn})\) from Little Group Constraints}\label{app:JOO-fc-derive}
Here we present a general method for solving the all-plus configuration of three-point correlators, with $\langle J^+O_2O_2\rangle$ as an illustrative example. Correlators of other helicities can subsequently be obtained by acting with the appropriate lowering operators.\\
The raising operator constraint forces the correlator to be independent of \(c_{11},~c_{22}\) and \(c_{33}\). 
Since the generators eq \eqref{AmbiGL1}, eq \eqref{AmbiSL2Diag} are constructed out of operators of the form \(c_{ij} \partial_{c_{ij}}\), it is convenient to introduce logarithmic variables \(l_{ij} = \log c_{ij}\) and define \(G_{ij} = \partial_{l_{ij}} \mathcal{G}\). In terms of these, the three \(GL(1,\R)\) eq \eqref{AmbiGL1} constraints become 
\begin{align}
    G_{31} - G_{13} + G_{21} - G_{12} = 0,~~G_{32} - G_{23} + G_{12} - G_{21}= 0,~~ G_{23} - G_{32} + G_{13} - G_{31} = 0, 
\end{align}
and the three diagonal constraints due to \(SL(2,\R)\) eq \eqref{AmbiSL2Diag} for $\langle J^+O_2O_2\rangle$ leads to the following constraints 
\begin{align}
    \notag G_{31} + G_{13} + G_{21}& + G_{12}  = -4\mathcal{G},~~G_{32} + G_{23} + G_{12} + G_{21}=  -2\mathcal{G},\\
    &G_{23} + G_{32} + G_{13} + G_{31} = -2\mathcal{G}.
\end{align}
Solving them simultaneously, we get
\begin{align}
    G_{12} = G_{31} = t,~~G_{21} = G_{13} = -t-2\mathcal{G},~~G_{23} = -G_{32} = t+\mathcal{G},
    \label{joo-equations}
\end{align}
where \(t(l_{ij})\) is some auxiliary function which cannot be fixed using only the \(GL(1,\mathbb{R})\) constraints and diagonal \(SL(2,\mathbb{R})\) constraints.
Each of the six first-order equations has the same form: a derivative of \(\mathcal{G}\) with respect to one \(l_{ij}\) is proportional to \(\mathcal{G}\), plus the same undetermined term \(t\). This is the multivariable analogue of \(\frac{dG}{dl}=aF+t\), to solve which we remove the homogeneous part by writing \(G=e^{al}H\), so that \(H\) now solves a simpler differential equation \(\frac{dH}{dl}=t\). We apply the same idea here. The coefficients of \(\mathcal G\) in the six equations determine an exponential prefactor. Factoring it out, we have
\begin{equation}
    \mathcal{G}=\exp(-2l_{21}-2l_{13}+l_{23}-l_{32})\,H(l_{12}, l_{21}, l_{13}, l_{31}, l_{23}, l_{32})
\end{equation}
Indeed, plugging in this ansatz, we see that 
\begin{align}
    \notag & G_{12} = \exp(-2l_{21}-2l_{13}+l_{23}-l_{32}) \frac{\partial H}{\partial l_{12}},~~~G_{21} = -2 \mathcal{G} + \exp(-2l_{21}-2l_{13}+l_{23}-l_{32})  \frac{\partial H}{\partial l_{21}}\\
    \notag & G_{13} = -2\mathcal{G} + \exp(-2l_{21}-2l_{13}+l_{23}-l_{32}) \frac{\partial H}{\partial l_{13}},~~~G_{31} =  \exp(-2l_{21}-2l_{13}+l_{23}-l_{32}) \frac{\partial H}{\partial l_{31}}\\
    & G_{23} = \mathcal{G} + \exp(-2l_{21}-2l_{13}+l_{23}-l_{32})   \frac{\partial H}{\partial l_{23}},~~~G_{32} = -\mathcal{G} + \exp(-2l_{21}-2l_{13}+l_{23}-l_{32})   \frac{\partial H}{\partial l_{32}}
\end{align}
Comparing this with equation eq \eqref{joo-equations}, we get the following set of constraints
\begin{align}
\frac{\partial H}{\partial l_{21}} = -&\frac{\partial H}{\partial l_{12}},
\quad \frac{\partial H}{\partial l_{13}} = -\frac{\partial H}{\partial l_{12}},\quad \frac{\partial H}{\partial l_{31}} = \frac{\partial H}{\partial l_{12}},\quad
\notag\frac{\partial H}{\partial l_{23}} = \frac{\partial H}{\partial l_{12}},\quad \frac{\partial H}{\partial l_{32}} = -\frac{\partial H}{\partial l_{12}}.
\end{align}
This set of constraints is satisfied if $H$ depends only on the combination
\begin{equation}
    u_l = l_{13}+l_{32}+l_{21} -l_{12}-l_{23}-l_{31}.
\end{equation}
The general ansatz, therefore, is 
\begin{equation}
    \mathcal{G} = \frac{c_{23}}{c_{21}^2 c_{13}^2 c_{32}} H\left( \frac{c_{21}c_{13}c_{32}}{c_{13}c_{32}c_{21}}   \right) = \frac{c_{23}c_{32}}{c_{21}^2 c_{13}^2 c_{32}^2} H\left(u\right) .
\end{equation}
With this general ansatz, demanding that the lowest helicity correlator be annihilated by the lowering operator, we obtain a differential equation purely in \(u\). For \(\braket{J^+OO}\), the differential equation is of the form 
\begin{equation}
    (u-1) u H'(u)+2 H(u) = 0\quad \implies  \quad H(u) = \frac{ u^2}{(u-1)^2}
\end{equation}
Therefore, we have 
\begin{equation}
    \mathcal{G}_{J^+O_2O_2}(c_{mn}) = \frac{c_{23} c_{32}}{\left(c_{12} c_{23} c_{31}-c_{13} c_{21} c_{32}\right){}^2}.
\end{equation}

   \section{Ambitwistor Scalar three-point integral}\label{app:AmbiDetail}
We define
\begin{equation}
S_{ij}\equiv Z_i\cdot W_j ,
\qquad
A\equiv c_{12}c_{23}c_{31}-c_{13}c_{21}c_{32}.
\end{equation}
 Thus $\langle O_2(Z_1,W_1)O_2(Z_2,W_2)O_2(Z_3,W_3)\rangle$ using eq \eqref{eq:Ambi-o2o2o2} is given by,
\begin{equation}
\mathcal I = \int \prod_{i,j=1}^{3}dc_{ij}\,
e^{i\sum_{i,j=1}^{3}c_{ij}S_{ij}}\,
\frac{1}{|A|}.
\end{equation}


The diagonal integrations give
\begin{equation}
\int dc_{11}dc_{22}dc_{33}\,
e^{i(c_{11}S_{11}+c_{22}S_{22}+c_{33}S_{33})}
=
\delta(S_{11})\delta(S_{22})\delta(S_{33}),
\end{equation}
Where overall Fourier \(2\pi\)-factors are absorbed into the measure.
Now, 
\begin{align}
    \mathcal{I} = \delta(S_{11})\delta(S_{22})\delta(S_{33}) \int \prod_{i\neq j=1}^3 dc_{ij} e^{i \sum_{i\neq j=1}^3c_{ij} S_{ij}} \frac{1}{|c_{23}c_{31}||\rho|},
\end{align}
where,
\begin{equation}
c_{12}^{\star}
=
\frac{c_{13}c_{21}c_{32}}{c_{23}c_{31}},
\qquad
\rho=c_{12}-c_{12}^{\star}.
\end{equation}
The \(c_{12}\)-dependent phase, after the change of variables, becomes
\begin{equation}
e^{ic_{12}S_{12}}
=
e^{ic_{12}^{\star}S_{12}}e^{i\rho S_{12}}.
\end{equation}
Now we use the finite-part prescription to solve the $\rho$ integral,
\begin{equation}
\operatorname{FP}\int_{\mathbb R}\frac{d\rho}{|\rho|}e^{i\rho S}
=
-2\log\left|{S}\right|.
\end{equation}
Therefore
\begin{equation}
\begin{aligned}
\mathcal I_{\rm sgn}
&=
-2\log\left|{S_{12}}\right|
\delta(S_{11})\delta(S_{22})\delta(S_{33})
\times
\int dc_{23}dc_{31}dc_{13}dc_{21}dc_{32}\,
\frac{e^{i\Phi_\star}}{|c_{23}c_{31}|},
\end{aligned}
\end{equation}
where,
\begin{equation}
\Phi_\star
=
c_{23}S_{23}
+c_{31}S_{31}
+c_{32}S_{32}
+c_{13}S_{13}
+c_{21}S_{21}
+
\frac{S_{12}c_{13}c_{21}c_{32}}{c_{23}c_{31}}.
\end{equation}
Next, doing two more integral of $c_{13}$ and $c_{21}$, we get
\begin{equation}
\begin{aligned}
\mathcal I
&=
-\frac{2}{|S_{12}|}
\log\left|\frac{S_{12}}{\mu}\right|
\delta(S_{11})\delta(S_{22})\delta(S_{33})
\\
&\quad\times
\int_{\mathbb R^3}
\frac{dc_{23}dc_{31}dc_{32}}{|c_{32}|}
\exp\left[
i(c_{23}S_{23}+c_{31}S_{31}+c_{32}S_{32})
-i\frac{S_{13}S_{21}}{S_{12}}
\frac{c_{23}c_{31}}{c_{32}}
\right].
\end{aligned}
\end{equation}
Now define
\begin{equation}
x=-c_{23}S_{23},
\qquad
y=-c_{31}S_{31},
\qquad
u=-c_{32}S_{32}.
\end{equation}
The reduced form is therefore
\begin{equation}
\boxed{
\begin{aligned}
\mathcal I
&= -2\frac{
\delta(S_{11})\delta(S_{22})\delta(S_{33})
}{
|S_{12}S_{23}S_{31}|
}
\log\left|{S_{12}}\right|
\times
\operatorname{FP}
\int_{\mathbb R^3}
\frac{dx\,dy\,du}{|u|}
\exp\left[
-i\left(
x+y+u-\tau\frac{xy}{u}
\right)
\right].
\end{aligned}
}
\end{equation}
If we evaluate the integrals in the order \(y\), \(u\) and \(x\), we get 
\begin{equation}
\operatorname{FP}
\int_{\mathbb R^3}
\frac{dx\,dy\,du}{|u|}
e^{-i(x+y+u-\tau xy/u)}
=
\delta(1+\tau).
\end{equation}
Thus, the scalar result is
\begin{equation}
\boxed{
\begin{aligned}
\mathcal I
&= -2\frac{
\delta(S_{11})\delta(S_{22})\delta(S_{33})
}{
|S_{12}S_{23}S_{31}|
}
\log\left|{S_{12}}\right|
\times
\delta(1+\tau).
\end{aligned}
}
\end{equation}
The \(\log|S_{12}| \) appeared as a consequence of doing the \(c_{12}\) integral first. This was one of the choices, and we could have chosen to start with any of the 6 variables. Each choice of \(c_{ij}\) gives the same result, but with \(\log|S_{ij}|\). To put the answer in a symmetric form, we add the results from all the choices, and also use the \(\delta(1+\tau)\) to obtain \(\log|S_{12}S_{23}S_{31}|\), which we have presented in the main text.


\section{Ambitwistor Penrose Transform for Spinning Correlators}\label{app:penrose-details}
In this section, we present the details of the Penrose transform to obtain position-space correlators. For illustrative purposes, let us consider the correlator \(\braket{J^+J^+O_2}\). \\
Under the Penrose transform, after dropping the diagonal terms, we have the expression 
\begin{align}
        \notag \braket{JJO} = \int dc_{ij} \frac{d^2\lambda_i d^2\tilde\lambda_i}{\mathrm{Vol}(GL(1, \mathbb R))^6} &(c_{21}\lambda_{2\alpha} + c_{31}\lambda_{3\alpha})(c_{12}\tilde\lambda_{2\dot\alpha} + c_{13}\tilde\lambda_{3\dot\alpha}) (c_{12}\lambda_{1\beta} + c_{32}\lambda_{3\beta})\\\notag&(c_{21}\tilde\lambda_{1\dot\beta} +  c_{23}\tilde\lambda_{3\dot\beta})~e^{i\sum_{i,j=1}^{3} c_{ij} \lambda_i \cdot x_{ij}\cdot \tilde\lambda_j} \frac{c_{13}c_{23}c_{31}c_{32}}{|c_{12}c_{23}c_{31}-c_{13}c_{21}c_{32}|^3}.
    \end{align}
    Notice that the rescaling redundancies of \(\lambda\) and \(\tilde \lambda\) are tied to the presence of \(c_{ij}\), in the sense that any rescaling of the spinors can be absorbed into \(c_{ij}\) without changing the integrand. Therefore, we do the following variable change that removes as many \(c_{ij}\)s as possible from the exponent
    \begin{equation}
        \lambda _1\to \frac{c_{32} \lambda _1}{c_{12}},~\lambda _2\to \frac{c_{31} \lambda _2}{c_{21}},~\lambda _3\to \lambda _3,~\tilde\lambda_1\to \frac{\tilde\lambda_1}{c_{31}},~\tilde\lambda_2\to \frac{\tilde\lambda_2}{c_{32}},~\tilde\lambda_3\to \frac{c_{12} \tilde\lambda_3}{c_{13} c_{32}},
    \end{equation}
    and we get the following expression
    \begin{align}
        \notag \int dc_{ij} \frac{d^2\lambda_i d^2\tilde\lambda_i}{\mathrm{Vol}(GL(1, \mathbb R))^6} (\lambda_{2\alpha} + \lambda_{3\alpha})(\tilde\lambda_{2\dot\alpha} + \tilde\lambda_{3\dot\alpha}) (\lambda_{1\beta} + \lambda_{3\beta})(c_{21}c_{13}c_{32}\tilde\lambda_{1\dot\beta} +  c_{12}c_{23}c_{31}\tilde\lambda_{3\dot\beta})~\\\frac{c_{12} c_{23} c_{31}}{c_{13}^2 c_{21}^2 c_{32}^2}\frac{1}{|c_{12}c_{23}c_{31}-c_{13}c_{21}c_{32}|^3} \exp\left(i\sum_{(i,j)\ne (2,3)} \lambda_i\cdot x_{ij}\cdot  \tilde\lambda_j + \frac{c_{12} c_{23} c_{31}}{c_{13} c_{21} c_{32}}\lambda_2\cdot x_{23}\tilde\cdot \lambda_3\right).
    \end{align}
    Now we do some trivial relabelings and one non-trivial change of variable
    \begin{equation}
        c_{12}\to a_2,~c_{13}\to a_3,~c_{21}\to a_4,~c_{31}\to a_5,~c_{32}\to a_6,~c_{23}\to \frac{a_1 \left(a_3 a_4 a_6\right)}{a_2 a_5},
    \end{equation}
    which puts the above expression in the following form
    \begin{align}
        \int d^2\lambda_i d^2\tilde\lambda_i da_i \frac{1}{|a_2||a_5|a_3^2 a_4^2a_6^2} \frac{a_1}{|a_1-1|^3 }
        \left(\lambda _{2\alpha} + \lambda_{3\alpha}\right) \left( \tilde\lambda_{2\dot\alpha}+ \tilde\lambda_{3\dot\alpha}\right) 
        \left(\lambda _{1\beta} +  \lambda_{3\beta}\right) \\\left( \tilde\lambda_{1\dot\beta}+ a_1\tilde\lambda_{3\dot\beta}\right) \exp\left(i\sum_{(i,j)\ne (2,3)} \lambda_i\cdot x_{ij}\cdot \tilde\lambda_j + a_1 \,\lambda_2\cdot x_{23}\cdot \tilde\lambda_3  \right),
    \end{align}
    where now all the \(a_i\) integrals except \(a_1\) are just volume integrals.\\
    By introducing the generating functional, we now evaluate the spinor integrals to obtain the position-space answer. Here we present the expressions for the first derivatives with respect to the source for the case of three points. Higher derivatives can be written in terms of the first derivative using the identity in eq \eqref{eq:source-n-derivative}
    \begin{align}
        \frac{\partial}{\partial S_{11}^{\alpha \dot\alpha}} \frac{1}{\det(X+S)}\Bigg|_{S=0} = -\frac{a_1 }{a_1-1}\left(\frac{x_{12 \alpha \dot{\alpha }}}{x_{12}^2}+\frac{x_{31 \alpha \dot{\alpha }}}{x_{13}^2}\right), \\
        \frac{\partial}{\partial S_{22}^{\alpha \dot\alpha}} \frac{1}{\det(X+S)}\Bigg|_{S=0} = -\frac{1}{a_1-1}\left( \frac{x_{12 \alpha \dot{\alpha }}}{x_{12}^2}+\frac{x_{23 \alpha \dot{\alpha }}}{x_{23}^2} \right),\\
        \frac{\partial}{\partial S_{33}^{\alpha \dot\alpha}} \frac{1}{\det(X+S)}\Bigg|_{S=0} = -\frac{1}{a_1-1}\left( \frac{x_{23 \alpha \dot{\alpha }}}{x_{23}^2}+\frac{x_{31 \alpha \dot{\alpha }}}{x_{13}^2} \right),\\
        \frac{1}{a_1}\,\frac{\partial}{\partial S_{12}^{\alpha \dot\alpha}} \frac{1}{\det(X+S)}\Bigg|_{S=0} = \frac{\partial}{\partial S_{21}^{\alpha \dot\alpha}} \frac{1}{\det(X+S)}\Bigg|_{S=0} = \frac{1}{\left(a_1-1\right)}\frac{x_{12 \alpha \dot{\alpha }}}{ x_{12}^2},\\
        \frac{\partial}{\partial S_{13}^{\alpha \dot\alpha}} \frac{1}{\det(X+S)}\Bigg|_{S=0} = \frac{1}{a_1}\,\frac{\partial}{\partial S_{31}^{\alpha \dot\alpha}} \frac{1}{\det(X+S)}\Bigg|_{S=0} = \frac{1}{\left(a_1-1\right)}\frac{x_{31 \alpha \dot{\alpha }}}{ x_{31}^2},\\
        \frac{\partial}{\partial S_{23}^{\alpha \dot\alpha}} \frac{1}{\det(X+S)}\Bigg|_{S=0} = \frac{\partial}{\partial S_{32}^{\alpha \dot\alpha}} \frac{1}{\det(X+S)}\Bigg|_{S=0} = \frac{1}{\left(a_1-1\right)}\frac{x_{23 \alpha \dot{\alpha }}}{ x_{23}^2}.
    \end{align}
    These can be put together appropriately to obtain the position space expression for \(\braket{JJO_2}\)

\section{Bulk Calculation of Chern-Simons Theory}\label{CST}
In this appendix, we present the details of the bulk computation for the three-point Witten diagram due to Chern-Simons interaction in AdS$_5$. The Chern-Simons term is metric-independent, and the relevant cubic interaction is given by
\begin{align}
S_{\rm int}
=
\kappa\int d^5x\;\Tr\left(A\wedge dA\wedge dA\right).
\end{align}
For the non-Abelian theory, the group-theory factor relevant for the
three-point correlator is the symmetric invariant
\begin{align}
d^{abc}=\Tr\!\left(T^{(a}T^bT^{c)}\right).
\end{align}
We work in radial gauge $A_z=0$, where the cubic interaction can be
written as
\begin{align}
S_{\rm int}
=
\frac{\kappa\;d^{abc}}{2}
\int dz\,d^4x\,
\epsilon^{\mu\nu\rho\sigma}
\Big[
A_\mu^a(\partial_z A_\nu^b)
(\partial_\rho A_\sigma^c-\partial_\sigma A_\rho^c)
+
A_\mu^a
(\partial_\rho A_\sigma^b-\partial_\sigma A_\rho^b)
(\partial_z A_\nu^c)
\Big].
\end{align}
Consequently, the Feynman rule for the vertex can be expressed compactly as
\begin{align}
V_{\mu\nu\rho}^{abc}
=
i\kappa\,d^{abc}\,
\epsilon_{\mu\nu\rho\sigma}
\left[
\partial_z^{(1)}(k_2-k_3)^\sigma
+\partial_z^{(2)}(k_3-k_1)^\sigma
+\partial_z^{(3)}(k_1-k_2)^\sigma
\right],
\end{align}
where $\partial_z^{(i)}$ acts only on the bulk-to-boundary propagator
associated with leg $i$. After contracting with the polarization vectors, we obtain the following expression for the vertex
\begin{align}
V
=
\epsilon(\epsilon_1,\epsilon_2,\epsilon_3,\mu)(k_2-k_3)^\mu
\,\partial_z^{(1)}
+
\epsilon(\epsilon_1,\epsilon_2,\epsilon_3,\mu)(k_3-k_1)^\mu
\,\partial_z^{(2)}
+
\epsilon(\epsilon_1,\epsilon_2,\epsilon_3,\mu)(k_1-k_2)^\mu
\,\partial_z^{(3)} .
\end{align}
In the radial gauge, the transverse bulk-boundary propagator for the gauge field $A_\mu$ is
\begin{equation}
K_\mu(k,z)
=
\epsilon_\mu(k)\,z k K_1(kz).
\end{equation}
Combining the vertex and the transverse propagators, the Witten diagram computation for the parity-odd three-current correlator is given by the following bulk integral
\begin{align}
\big(\epsilon(\epsilon_1,\epsilon_2,\epsilon_3,\mu)(k_2-k_3)^\mu
\,\partial_z^{(1)}
+\textrm{cyclic}\big)\int_0^\infty dz
\left(zk_1K_1(k_1z)\right)
\left(zk_2K_1(k_2z)\right)
\left(zk_3K_1(k_3z)\right).
\end{align}
To obtain the Wightman function, it is necessary to take discontinuities with respect to leg 1 and leg 3 \cite{Bala:2026trw}. This amounts to replacing the Bessel $K$ functions on legs 1 and 3 with the corresponding Bessel $J$ representation. The required $JKJ$ master integral can be
generated from
\begin{align}
I_{1\{000\}}
=
\frac{1}{\sqrt{J_2}},
\end{align}
where $J_2=-(k_1+k_2+k_3)(k_1+k_2-k_3)(k_1-k_2+k_3)(-k_1+k_2+k_3).$ The higher-index integrals needed after radial differentiation can be
generated through
\begin{align}
I_{3\{110\}}
=
k_1k_2
\frac{\partial^2 I_{1\{000\}}}
{\partial k_1\partial k_2},
\qquad
I_{3\{101\}}
=
k_1k_3
\frac{\partial^2 I_{1\{000\}}}
{\partial k_1\partial k_3},
\qquad
I_{3\{011\}}
=
k_2k_3
\frac{\partial^2 I_{1\{000\}}}
{\partial k_2\partial k_3}.
\end{align}
Using these properties of the Bessel functions, the correlator can then be written as 
\begin{align}
\braket{ J_1J_2J_3}_{\rm odd}
\propto
d^{abc}
\big(&
\epsilon(\epsilon_1,\epsilon_2,\epsilon_3,\mu)(k_3-k_1)^\mu\,I_{3\{101\}}
+
\epsilon(\epsilon_1,\epsilon_2,\epsilon_3,\mu)(k_1-k_2)^\mu\,I_{3\{110\}}\notag\\
+&
\epsilon(\epsilon_1,\epsilon_2,\epsilon_3,\mu)(k_2-k_3)^\mu\,I_{3\{011\}}
\big),
\end{align}
Using momentum conservation and epsilon-tensor identities, the result
can be reorganized into the compact form
\begin{equation}
\braket{J_1J_2J_3}_{\rm odd}
\propto
d^{abc}
\frac{k_1^2k_2^2k_3^2}{J_2^{5/2}}
\Big[
\epsilon(\epsilon_1,\epsilon_2,\epsilon_3,k_1)
(k_1^2-k_2^2-k_3^2)(k_2^2-k_3^2)
+\text{cyclic}
\Big].
\end{equation}
This expression has the expected parity-odd structure and
satisfies the special conformal Ward identity when the Wightman
prescription is implemented consistently. Moreover, we see that this expression matches the answer obtained via the Grassmannian eq \eqref{JJJOdd}.

\bibliography{biblio}
\bibliographystyle{JHEP} 

\end{document}